\documentclass[conference]{IEEEtran}
\usepackage{times}

\usepackage[numbers]{natbib}
\usepackage{multicol}
\usepackage[bookmarks=true]{hyperref}
\usepackage[dvipsnames]{xcolor}
\usepackage{graphicx}
\usepackage{multicol}
\usepackage{caption}
\usepackage{amsmath}
\usepackage{cleveref}
\usepackage{amsthm}
\usepackage{subcaption}
\usepackage{mathtools}
\usepackage{booktabs}
\usepackage{tikz}
\hypersetup{
    colorlinks=true,
    linkcolor=blue,
    filecolor=magenta,      
    urlcolor=blue,
    pdfpagemode=FullScreen,
    }

\usepackage{amssymb}
\usepackage{bm}
\usepackage{stmaryrd}
\usepackage{cancel}

\newcommand{\T}{^\mathrm{T}}
\newcommand{\dtot}{\mathrm{d}}

\DeclareMathOperator*{\argmin}{arg\,min}
\DeclareMathOperator*{\tr}{tr}

\newcommand{\diag}{\mathrm{diag}}
\newcommand{\blkdiag}{\mathrm{blkdiag}}

\newcommand{\va}{{\bf a}}

\newcommand{\vc}{{\bf c}}
\newcommand{\vd}{{\bf d}}

\newcommand{\vf}{{\bf f}}
\newcommand{\vg}{{\bf g}}

\newcommand{\vq}{{\bf q}}

\newcommand{\vu}{{\bf u}}

\newcommand{\vw}{{\bf w}}
\newcommand{\vx}{{\bf x}}
\newcommand{\vy}{{\bf y}}
\newcommand{\vz}{{\bf z}}

\newcommand{\vA}{{\bf A}}

\newcommand{\vC}{{\bf C}}

\newcommand{\vG}{{\bf G}}

\newcommand{\vI}{{\bf I}}

\newcommand{\vM}{{\bf M}}

\newcommand{\vR}{{\bf R}}

\newcommand{\vX}{{\bf X}}

\newcommand{\Rb}{\mathbb{R}}

\newcommand{\Zb}{\mathbb{Z}}

\newcommand{\calC}{\mathcal{C}}

\newcommand{\calH}{\mathcal{H}}
\newcommand{\calI}{\mathcal{I}}
\newcommand{\calL}{\mathcal{L}}
\newcommand{\calN}{\mathcal{N}}
\newcommand{\calP}{\mathcal{P}}

\newcommand{\vSigma}{\mathbf{\Sigma}}

\newcommand{\bdelta}{\bm{\delta}}
\newcommand{\bepsilon}{\bm{\epsilon}}
\newcommand{\bzeta}{\bm{\zeta}}
\newcommand{\blambda}{\bm{\lambda}}
\newcommand{\bnu}{\bm{\nu}}
\newcommand{\bchi}{\bm{\chi}}
\newcommand{\bgamma}{\bm{\gamma}}

\newcommand{\bomega}{\bm{\omega}}

\usepackage{tikz}
\usetikzlibrary{shapes.geometric,backgrounds,calc,positioning}

\definecolor{mat_red}{HTML}{B71C1C}
\definecolor{mat_green}{HTML}{33691E}
\definecolor{mat_blue}{HTML}{01579B}
\definecolor{mat_grey}{HTML}{ECEFF1}
\definecolor{mat_dark_grey}{HTML}{546E7A}
\tikzset{
  mybackground/.style={execute at end picture={
        \begin{scope}[on background layer]
          \draw[mat_dark_grey,fill=mat_grey,rounded corners=1ex,line width=0.01pt] ($(current bounding box.south west)+(-.5,-.5)$)
                    rectangle ($(current bounding box.north east)+(.5,.5)$);
        \end{scope}
    }},
}

\newcommand*\blambdadup{\blambda_{\text{dup}}}

\newcommand*\blambdadupactive{\blambda_{\text{dup}, \text{active}}}
\newcommand*\blambdadupinactive{\blambda_{\text{dup}, \text{inactive}}}

\newcommand*\blambdanondup{\blambda_{\text{non-dup}}}
\newcommand*\blambdanondupactive{\blambda_{\text{non-dup}, \text{active}}}
\newcommand*\blambdanondupinactive{\blambda_{\text{non-dup}, \text{inactive}}}

\newcommand*\vcdij{\vc_{\zeta_{l,j}}}
\newcommand*\vcdi{\vc_{\zeta_{l}}}

\newcommand{\lambdadup}[1]{\lambda_{\text{dup}, {#1}}}
\newcommand{\lambdanondup}[1]{\lambda_{\text{non-dup}, {#1}}}

\newcommand{\lamdlamdup}[1]{\lambda_{\text{dup}, {#1}} \, \dif \lambda_{\text{dup}, {#1}}}
\newcommand{\lamdlamnondup}[1]{\lambda_{\text{non-dup}, {#1}} \, \dif \lambda_{\text{non-dup}, {#1}}}

\newcommand*\dif{\mathop{}\!\mathrm{d}}

\newcommand*\dvu{\dif \vu}
\newcommand*\dvR{\dif \vR}
\newcommand*\dvC{\dif \vC}

\newcommand*\dblambdadup{\dif \blambdadup}

\newcommand*\dblambdadupactive{\dif \blambdadupactive}

\newcommand*\dblambdanondup{\dif \blambdanondup}
\newcommand*\dblambdanondupactive{\dif \blambdanondupactive}

\newcommand*\df{\dif f}
\newcommand*\dvX{\dif \vX}

\newcommand*\dbepsilon{\dif \bepsilon}
\newcommand*\dell{\dif \ell}

\newcommand*\vzero{\bf 0}

\newcommand*\cactive{{\text{active}}}
\newcommand*\cinactive{{\text{inactive}}}

\newcommand*\Nineqi{{N_{\text{ineq},i}}}
\newcommand*\Nineqb{{\bar{N}_\text{ineq}}}

\newcommand*\IP{{\textrm{IP}}}

\newtheorem{assumption}{Assumption}
\crefname{assumption}{assumption}{assumptions}

\newtheorem{innercustomgeneric}{\customgenericname}  
\providecommand{\customgenericname}{}
\newcommand{\newcustomtheorem}[2]{%
  \crefname{#2}{#2}{#2s}%
  \newenvironment{#1}[1]
  {%
   \renewcommand\customgenericname{#2}%
   \crefalias{innercustomgeneric}{#2}%
   \renewcommand\theinnercustomgeneric{##1}%
   \innercustomgeneric
  }
  {\endinnercustomgeneric}
}
\newcustomtheorem{customthm}{Theorem}
\newcustomtheorem{customlemma}{Lemma}

\begin{document}

% paper title
\title{Decentralized Safe Multi-agent Stochastic Optimal Control using Deep FBSDEs and ADMM}

% You will get a Paper-ID when submitting a pdf file to the conference system
%\author{Author Names Omitted for Anonymous Review. Paper-ID [232]}

% \author{\authorblockN{Michael Shell\authorrefmark{1}}
% \thanks{\authorrefmark{1}equal contribution}
% \authorblockA{School of Electrical and\\Computer Engineering\\
% Georgia Institute of Technology\\
% Atlanta, Georgia 30332--0250\\
% Email: mshell@ece.gatech.edu}
% \and
% \authorblockN{Homer Simpson\authorrefmark{1}}
% \authorblockA{Twentieth Century Fox\\
% Springfield, USA\\
% Email: homer@thesimpsons.com}
% \and
% \authorblockN{James Kirk\\ and Montgomery Scott}
% \authorblockA{Starfleet Academy\\
% San Francisco, California 96678-2391\\
% Telephone: (800) 555--1212\\
% Fax: (888) 555--1212}}

% avoiding spaces at the end of the author lines is not a problem with
% conference papers because we don't use \thanks or \IEEEmembership

% for over three affiliations, or if they all won't fit within the width
% of the page, use this alternative format:
% 
\author{\authorblockN{Marcus A. Pereira\authorrefmark{1}\authorrefmark{2},
Augustinos D. Saravanos\authorrefmark{1}\authorrefmark{2},
Oswin So\authorrefmark{3} and 
Evangelos A. Theodorou\authorrefmark{1}}
\authorblockA{\authorrefmark{1}Daniel Guggenheim School of Aerospace Engineering, Georgia Institute of Technology, Atlanta, GA}
\authorblockA{\authorrefmark{3}College of Computing, Georgia Institute of Technology, Atlanta, GA}
% \authorblockA{\{mpereira30@gatech.edu, asaravanos@gatech.edu, oswinso@gatech.edu, evangelos.theodorou@gatech.edu\}}
\authorblockA{\{mpereira30, asaravanos, oswinso, evangelos.theodorou\}@gatech.edu}
\authorblockA{\authorrefmark{2}These authors contributed equally}}

\maketitle

\begin{abstract}
In this work, we propose a novel safe and scalable decentralized solution for multi-agent control in the presence of stochastic disturbances. Safety is mathematically encoded using stochastic control barrier functions and safe controls are computed by solving quadratic programs. Decentralization is achieved by augmenting to each agent's optimization variables, copy variables, for its neighbors. This allows us to decouple the centralized multi-agent optimization problem. However, to ensure safety, neighboring agents must agree on \textit{what is safe for both of us}, creating a need for consensus. To enable safe consensus solutions, we incorporate an ADMM-based approach. Specifically, we propose a Merged Consensus ADMM-OSQP implicit neural network layer, that solves a mini-batch of both, local quadratic programs as well as the overall consensus problem, as a single optimization problem. This layer is embedded within a Deep Forward-Backward Stochastic Differential Equations (FBSDEs) network architecture at every time step, to facilitate end-to-end differentiable, safe and decentralized stochastic optimal control. The efficacy of the proposed approach is demonstrated on several challenging multi-robot tasks in simulation. By imposing collision avoidance constraints, the safe operation of all agents is ensured during the entire training process. We also demonstrate superior scalability in terms of computational and memory savings as compared to a centralized approach.
\end{abstract}

\IEEEpeerreviewmaketitle

\section{Introduction}
A vast variety of robotics applications such as coverage control \cite{cortes2004coverage}, flocking of UAVs \cite{olfati2006flocking}, multi-robot navigation \cite{alonso2016distributed}, etc., falls into the class of multi-agent control problems. Such settings usually include a team of autonomous agents which are required to cooperate in order to accomplish a common goal. Effective frameworks for addressing these problems should be able to control the agents in an \textit{optimal} manner, while ensuring their \textit{safety under uncertainty}. In addition, it is of paramount importance that such methods are \textit{scalable} to large-scale systems in terms of computational efficiency, memory usage and communication requirements.

% \augustinos{Optimal control methods that rely on distributed optimization have been gaining significant attention in tackling these problems, since they can yield elegant decentralized solutions \cite{tang2021fast, xiao2019decentralized, halsted2021survey}. Furthermore, recently proposed frameworks that combine stochastic control and distributed optimization principles \cite{saravanos2021distributed, rostampour2019distributed} have shown to be capable of successfully encompassing both safety and scalability.}

\par Deep reinforcement learning has enjoyed significant fame over the past few years \cite{mnih2013playing, silver2017mastering, silver2018general, lillicrap2015continuous}, although being restricted to simulation settings where safety is not a primary concern. Therefore, very recently, the focus has shifted to the area of safe reinforcement learning \cite{wang2018safe, cheng2019end}, where safety is required during the entire training process.
\par Control Barrier Functions (CBFs) \cite{ames2016control, ames2019control}, have been popularly used in the robotics community to design controllers that guarantee invariance of user-defined safe sets. These are generally combined with off-the-shelf Quadratic Programming (QP) solvers to deliver real-time safe control solutions. However, most work has focused primarily on deterministic systems. Very recent works employing, so called, Stochastic CBFs (SCBFs) \cite{clark2021control, santoyo2021barrier, sarkar2020high, yaghoubi2020risk} aim to bridge the gap, but lack scalability to large scale systems.
\par The confluence of Deep Learning and traditional optimization methods for \textit{intra-layer optimization} has been of special interest lately. In particular, recent works embed QPs \cite{agrawal2019differentiable, amos2017optnet, amos2018differentiable}, root-finding methods \cite{bai2019deep} and even non-convex solvers \cite{amos2020differentiable, exarchos2020novas, finn2017model} into the forward-pass of deep neural network layers and are popularly referred to as \textit{implicit} neural network layers (other examples are \cite{chen2018neural, li2020scalable}). In most of these methods, the backward-pass is efficiently computed by invoking the implicit function theorem rather backpropagating through the unrolled graph of the forward-pass. 
\par A general approach for continuous-time Stochastic Optimal Control (SOC) relying on Forward-Backward Stochastic Differential Equations (FBSDEs) was recently combined with deep learning \cite{han2018solving} to solve high-dimensional Hamilton-Jacobi-Bellman PDEs (HJB-PDEs). Most noteworthy being deep FBSDEs \cite{pereira2019learning, pereira2020feynman, wang2019deep, chen2021large}, a scalable framework for SOC problems that, at its core, leverages the function approximation capabilities of deep recurrent neural networks (specifically Long Short-Term Memory networks or LSTMs) to learn the gradient of the value-function, which can then be used to compute optimal control policies. 

% Distributed optimization-based frameworks have been gaining significant attention for tackling multi-agent control problems. A suitable method for deriving such algorithms is the Alternating Direction Method of Multipliers (ADMM) \cite{boyd2011distributed}. In particular, ADMM-based methods have been recently proposed \cite{tang2021fast, xiao2019decentralized, halsted2021survey, cheng2021admm}, yielding elegant decentralized solutions for multi-agent control. Furthermore, recent works employing ADMM in a stochastic setting, \cite{saravanos2021distributed, rostampour2019distributed, le2020gaussian}, have shown to be capable of successfully encompassing both the safety under uncertainty and scalability desired attributes.
%
Distributed optimization-based frameworks have been gaining significant attention for tackling multi-agent control problems recently. A suitable method for deriving such algorithms is the Alternating Direction Method of Multipliers (ADMM) \cite{boyd2011distributed}. In particular, several ADMM-based methods such as \cite{cheng2021admm, halsted2021survey, tang2021fast}, have been proposed, yielding elegant decentralized solutions for multi-agent control. Furthermore, recent works employing ADMM in a stochastic setting \cite{saravanos2021distributed, le2020gaussian}, have shown to be capable of successfully encompassing both the safety under uncertainty and scalability desired attributes.

\par There has been very little work trying to put together all these ingredients into one framework.
The Safe Deep FBSDEs framework \cite{pereira2020safe} is one recent approach that partly addresses this need, the missing component being decentralization. More specifically, it uses an instantiation of implicit layers to solve embedded SCBF-based QPs combined with deep learning. Additionally, it utilizes a SCBF that tries to gaurantee safety with a probability of 1. The method was tested on low-dimensional systems in simulation and the resulting policies appeared conservative. Inspired by Safe Deep FBSDEs with the aim to overcome its limitations in order to scale to large multi-agent SOC problems, we propose a decentralized approach to safety. To this end, the specific contributions of our work are,
\begin{enumerate}
    \item A safe end-to-end differentiable framework that uses a novel SCBF formulation that is practically more meaningful and allows for safe, yet non-conservative policies,
    \item A \textit{fully decentralized} ADMM-based algorithm for large scale QPs embedded into an implicit neural network layer with drastic improvements in memory efficiency and training time as compared to a centralized approach, 
    \item A safe reinforcement framework, i.e., one that ensures safety not only at test time, but also while training the policy, and, 
    \item Extensive testing on several multi-robot challenging tasks in simulation, demonstrating the capabilities of the proposed framework.
\end{enumerate}

\section{Notation}
Here, we introduce the notation followed throughout this paper. Non-bold symbols are used for scalars $a \in \Rb$, and bold lowercase and uppercase symbols for vectors $\va \in \Rb^n$ and matrices $\vA \in \Rb^{n \times m}$, respectively. With $\vA[i,j]$, we refer to the element of the $i$-th row and $j$-th column of $\vA$. In addition, with $\va = [\va_1 ; \dots ; \va_N]$, we denote the vertical concatenation of vectors $\va_1, \dots, \va_N$. The $\ell_2$-norm of a vector $\va \in \Rb^n$ is defined as $\| \va \|_2 = \sqrt{\sum_{i=1}^n a_i^2}$ where $\va = [a_1 ; \dots ; a_N]$. Moreover, the trace of a matrix $\vA \in \Rb^{n \times n}$ is denoted with $\tr(\vA)$. We also define as $\mathrm{diag}(a_1, \dots, a_N) \in \Rb^{N \times N}$ the diagonal matrix with diagonal elements scalars $a_1, \dots, a_N$, and as $\mathrm{bdiag}(\vA_1, \dots, \vA_N)$ the block diagonal matrix constructed by the matrices $\vA_1, \dots, \vA_N$. Given a set $\calN$, its cardinality is denoted by $|\calN|$. With $\llbracket a, b\rrbracket$, we denote the integer interval $[a,b] \cap \Zb$. Finally, given a convex set $\calC$, $\Pi_{\calC}(\vx)$ denotes the projection of a vector $\vx$ onto the set and $\calI_c(\vx)$ denotes the set indicator function such that $\calI_c(\vx) = 0$ if $\vx \in \calC$ and $\calI_c(\vx) = + \infty$ otherwise.

% \textcolor{red}{\begin{enumerate}
%     \item symbols for scalar, vectors and matrices
%     \item semi-colon
%     \item integer sets
%     \item l2 norm 
%     \item set projection 
%     \item indicator function 
%     \item set cardinality 
% \end{enumerate}}

\section{Problem Formulation}
%
% \textcolor{red}{Let us consider a network of $N$ agents defined by the graph $\calG(\calV, \calE)$, where $\calV = \{1, \dots, N\}$ is the node set containing the agents and $\calE$ is the edge set indicating which agents are interconnected. The neighbors of each agent $i$ are provided by the set $\calN_i = \{ j \in \calV \ | \ j \neq i, \ (i,j) \in \calE \}$. (\textbf{We might want to change to a simpler non-graphical description)} }
The framework developed in this paper can be applied to various problems, however, we focus on a multi-robot setting so as to establish a notion of distance as well as to provide context for constraints commonly encountered in this setting.\par Consider a collection of $N$ agents. The stochastic, nonlinear and control-affine dynamics for any agent $i \in \llbracket 1, N \rrbracket$ are given by the following stochastic differential equation, \begin{equation}
\dtot \vx_i = \big( \vf_i (\vx_i) + \vG_i(\vx_i) \vu_i \big) \dtot t + \vSigma_i(\vx_i) \dtot \vw_i
\end{equation} where $\vx_i = \vx_i(t) \in \Rb^{n_i}$, $\vu_i = \vu_i(t) \in \Rb^{m_i}$, $\vw_i = \vw_i(t) \in \Rb^{p_i}$ are the state, control and standard Brownian-motion vectors, respectively, and $\vf_i(\cdot):\mathbb{R}^{n_i} \rightarrow \mathbb{R}^{n_i}$, $\vG_i(\cdot):\mathbb{R}^{n_i} \rightarrow \mathbb{R}^{n_i \times m_i}$ and $\vSigma_i(\cdot):\mathbb{R}^{n_i} \rightarrow \mathbb{R}^{n_i \times p_i}$ denote the drift vector, actuation matrix and diffusion matrix, respectively. In order to state the centralized problem, we construct the global state, control and Brownian-motion vectors, $\vx=[\vx_1;\,\vx_2;\,\ldots;\,\vx_N]$, $\vu=[\vu_1;\,\vu_2;\,\ldots;\,\vu_N]$, and $\vw=[\vw_1;\,\vw_2;\,\ldots;\,\vw_N]$ by concatenation. Thus, the global stochastic dynamics are provided as follows,
\begin{equation}
\dtot \vx = \big( \vf (\vx) + \vG(\vx) \vu \big) \dtot t + \vSigma(\vx) \dtot \vw
\label{eqn:global_dyn}
\end{equation}
with $\vf=[\vf_1;\,\vf_2;\,\ldots;\,\vf_N],\,\vG=\mathrm{bdiag}(\vG_1, \dots, \vG_N)$ and $\vSigma=\mathrm{bdiag}(\vSigma_1, \dots, \vSigma_N)$.
\par Stochastic Optimal Control (SOC) aims to minimize an expected cost subject to \eqref{eqn:global_dyn} given by, \begin{align}
    &\mathcal{J}(\vx,\,\vu,\,t_0)=\sum_{i=1}^N J_i(\vx_i,\,\vu_i,\,t_0) \nonumber\\ 
    &\,\,\,= \sum_{i=1}^N \mathbb{E}\bigg[\phi_i\big(\vx_i(T)\big) + \int_{t_0}^T \Big(c_i\big(\vx_i\big) + \frac{1}{2}\vu_i\T\vR_i\vu_i\Big)\,\dtot t  \bigg]\label{eqn:total_cost}
\end{align}
where for any agent $i$, $\phi_i(\cdot)$ is the terminal state-cost function and the running cost comprises of a purely state-dependent term $c_i(\cdot)$, and a quadratic control-cost term with weighting coefficients given by the matrix $\vR_i \in \mathbb{R}^{m_i \times m_i}$.  
\par To solve the SOC problem using dynamic programming, we define the value function as $V(\vx,\,t)=\inf_{\vu}\mathcal{J}(\vx,\,\vu,\,t)$, i.e., the optimal \textit{cost-to-go} from state $\vx$ at time step $t$. Next, using Ito's formula \cite[Chapter 4]{shreve2004stochastic}, one can derive the Hamilton-Jacobi-Bellman Partial Differential Equation (HJB-PDE), \begin{align}
    \frac{\partial V}{\partial t} + \inf_\vu \mathcal{H} = 0,\, V(\vx,\,T)&=\sum_{i=1}^N\phi_i\big(\vx_i(T)\big)
    \label{eqn:centralized_hjb_pde}\\ 
    \text{where, }\mathcal{H}= \frac{1}{2} \tr \bigg(\frac{\partial^2 V}{\partial \vx^2}\vSigma\vSigma\T\bigg)+&\frac{\partial V}{\partial \vx}\T\bigg(\vf+\vG\vu\bigg)\nonumber\\
    &+ \sum_{i=1}^N\Big(c_i+\frac{1}{2}\vu_i\T\vR_i\vu_i\Big)\nonumber
\end{align}
which is a backward semilinear PDE and $\mathcal{H}$ is referred to as the Hamiltonian. Owing to the construction of the global state and control by concatenation, the Hamiltonian minimization can be split into a sum of decoupled minimizations of the Hamiltonians associated with each agent $i\,\big(\text{i.e., }\inf_\vu \mathcal{H} = \sum_{i=1}^N \inf_{\vu_i} \mathcal{H}_i\big)$ in the absence of inter-agent safety constraints.
However, when the latter are considered, this decoupling is no longer valid. In this work, we adopt a probabilistic approach to safety by imposing such constraints using Stochastic Control Barrier Functions (SCBFs). We assume that a safe set defined by $\mathcal{S}=\{\vx:\,h(\vx)\geq 0\}$ is known, where $h(\vx)$ is task-dependent and usually hand-designed and the goal is to stay within the safe set (with a high probability) for the entire time horizon. In the context of multi-robot systems, these SCBFs can encode obstacle and inter-agent collision avoidance constraints.

There are two types of SCBFs currently in literature -- \textit{almost-sure} (type-I) and those which allow for violations (type-II). The type-I SCBFs ensure that the system remains inside the safe set with a probability of 1 \cite{clark2021control} for all time $t\in[0,\,T]$. These were successfully utilized within a recent deep-learning based SOC framework \cite{pereira2020safe}, but simulations suggest that the framework leads to conservative policies preventing agents from getting close to each other, thus, reducing the flexibility of the trained policy. This restriction can prohibit application to systems with a large number of agents. On the other hand, type-II SCBFs allow tuning the probability of failure based on one's risk appetite and therefore have a higher practical appeal. We hypothesize that policies trained using Type-II SCBFs would result in less conservative and therefore more scalable policies. These have been employed in recent works \cite{santoyo2021barrier, yaghoubi2020risk} and are based on derivation of Lyapunov functions for finite time stability of stochastic systems \cite[Chapter 3]{kushner1967stochastic}. Restating the result from \cite[Proposition 1]{santoyo2021barrier} here for convenience of the reader: \textit{suppose there exists a twice differentiable function $B(\vx)$, that satisfies the following inequalities, \begin{align}
    &B(\vx)  \geq 0 \quad \forall\,\vx \in \mathbb{R}^{Nn_i} \label{eqn:ineq_1}\\
    &B(\vx) \geq 1 \quad\forall\, \vx \in (\mathbb{R}^{Nn_i}\backslash \mathcal{S})\label{eqn:ineq_2}\\
    &\frac{\partial B}{\partial \vx}\T\Big(\vf + \vG \vu\Big) + \frac{1}{2}\text{tr}\bigg(\frac{\partial^2 B}{\partial \vx^2}\vSigma\vSigma\T\bigg) \leq -\alpha B(\vx) + \beta, \label{eqn:safety_constraint}
\end{align}
where \eqref{eqn:safety_constraint} is satisfied $\forall \,t\in[0,\,T]$ and $\forall\,\vx\in \mathbb{R}^{Nn_i}$ for some $\alpha \geq 0$ and $\beta \geq 0$. Based on the chosen values of $\alpha$ and $\beta$, one can compute bounds on the probability of failure. We refer the reader to \Cref{sec:failure_bounds_section} of the supplementary material for additional details.} 
\par Inequalities \eqref{eqn:ineq_1} and \eqref{eqn:ineq_2} can be satisfied by choosing $B(\vx)=e^{-\gamma h(\vx)}$ so that when the system exits the safe set $\mathcal{S}$, then $B(\vx)>1$ because $h(\vx)<0$. To satisfy \eqref{eqn:safety_constraint}, we impose it as a hard constraint on the Hamiltonian minimization in \eqref{eqn:centralized_hjb_pde}. Since the objective $\calH(\vu)$ is quadratic and the constraint \eqref{eqn:safety_constraint} is linear in $\vu$, a safe optimal control can be obtained by solving the resulting Quadratic Program (QP) at every time step. Similar to the work in \cite{pereira2020safe}, we have the following HJB-PDE, \begin{align}
&\frac{\partial V}{\partial t} + \inf_{\vu\in\vu_\text{safe}} \Bigg\{ \sum_{i=1}^N \bigg( \frac{1}{2}\vu_i\T\vR_i\vu_i + \frac{\partial V}{\partial \vx_i}\T \vG_i\vu_i \bigg) \Bigg\}  \nonumber\\&\qquad\qquad+\sum_{i=1}^N \bigg(c_i + \frac{\partial V}{\partial \vx_i}\T\vf_i + \frac{1}{2} \tr \Big(\frac{\partial^2 V}{\partial \vx_i^2}\vSigma_i\vSigma_i\T\Big)\bigg)=0\label{eqn:constrained_centralized_HJB}\\
    &\text{where, }\vu_\text{safe}= \Bigg\{\vu\Bigg|\frac{\partial B_k}{\partial \bar{\vx}_k}\T\bigg(\bar{\vf}_k+\bar{\vG}_k\bar{\vu}_k\bigg) \nonumber\\
    &+\frac{1}{2} \tr \bigg(\frac{\partial^2 B_k}{\partial \bar{\vx}_k^2}\bar{\vSigma}_k\bar{\vSigma}_k\T\bigg)\leq -\alpha B_k(\bar{\vx}_k) + \beta,\, \forall\,k\in \llbracket 1,N_\text{ineq} \rrbracket \Bigg\}\nonumber
\end{align}
where each $B_k$ is either a pairwise safety constraint between two agents or between an agent and an obstacle. The vectors $\bar{\vx}_k\text{ and }\bar{\vu}_k$ are obtained by concatenating states and controls of the two agents corresponding to $B_k$ for inter-agent constraints or just the state and the control vectors of the ego agent for agent-obstacle constraints. Finally, $\bar{\vf}_k,\,\bar{\vG}_k\text{ and }\bar{\vSigma}_k $ are constructed similar to $\bar{\vx}_k$ and $N_\text{ineq}=$$ N\choose2$$+NN_o$, is the total number of inequality constraints. The $N\choose2$ constraints account for all possible agent pairs for collision avoidance similar to work in \cite[Section 4.3.2]{pereira2020safe} and $N_o$ is the number of obstacles.  However, this clearly would not scale for large values of $N$. Further, due to the inter-agent constraints, the minimization of individual $\mathcal{H}_i$ can no longer be decoupled. Thus, the safe optimal control $\vu^*$, has to be solved as one large QP, i.e, in a \textit{centralized} manner.

\section{Centralized Solution Using Deep FBSDEs}
Before we propose our decentralized solution to address scalability, we summarize the deep learning based solution to safe SOC problems adapted from recent work \cite{pereira2020safe}, which if applied directly to solve \Cref{eqn:constrained_centralized_HJB} would lead to a centralized approach. 
\par The unique solution of \eqref{eqn:constrained_centralized_HJB} is linked to that of a system of FBSDEs via the Nonlinear Feynman-Kac lemma (we refer the reader to \Cref{sec:nonlinear_feynman_kac_derivation_section} of the supplementary material for the derivation). Assuming that a solution $\vu^*$ exists, the FBSDE system that solves \eqref{eqn:constrained_centralized_HJB} is given by, 
\begin{align}
    &\textbf{(FSDE)}\quad\dtot \vx = (\vf + \vG\vu^*)\dtot t + \vSigma \dtot\vw,\, \vx(0)=\vx_0\label{eqn:fsde}\\
    &\textbf{(BSDE)}\begin{cases}
    \dtot V = -\bigg[\sum_{i=1}^N\Big(c_i+\frac{1}{2}\vu_i^*{}\T\vR_i\vu_i^*\Big)\bigg]\dtot t + \frac{\partial V}{\partial \vx}\T \vSigma \dtot\vw \\
    V\big(\vx(T)\big) = \sum_{i=1}^N \phi_i\big(\vx_i(T)\big)\label{eqn:bsde}
    \end{cases}\\
    &\text{where, }\vu^* = \argmin_{\vu \in \vu_\text{safe}} \mathcal{H}. \nonumber
\end{align} Deep FBSDEs use deep neural networks to learn $\frac{\partial V}{\partial \vx}(\vx,\,t;\,\theta)$, which can be used to compute optimal control policies $\vu^*(\vx)$. Traditional methods \cite{exarchos2018stochastic, exarchos2019stochastic, exarchos2018stochasticL1, exarchos2016game} to solve FBSDEs relied on back-propagating and approximating the conditional expectation of the value function, $\mathbb{E}\big[V(\vx,\,t)\big]$, using least squares. This approach is prone to numerical ill-conditioning issues depending on which area of the state-space the system visits, requires hand-picking basis functions and suffers from compounding least squares errors over time as $\mathbb{E}\big[V(\vx,\,t)\big]$ is back-propagated from $t=T$ to $t=t_0$. Deep FBSDEs circumvent the need to back-propagate $\mathbb{E}\big[V(\vx,\,t)\big]$ by instead parameterizing $\hat{V}\big(\vx(0),\,0\big)$ with trainable weights. Using this approximation of the initial condition, $V(\vx,\,t)$ can then be forward propagated similar to a forward SDE using \eqref{eqn:bsde}. At the end of the time horizon, the predicted terminal value $\hat{V}\big(\vx(T),\,T\big)$ that relies on the predictions of $\frac{\partial V}{\partial \vx}(\vx,\,t;\,\theta)$ provided by a deep LSTM network, is compared to the true terminal value $V\big(\vx(T),\,T\big)$ to construct a loss function. $V\big(\vx(T),\,T\big)$ is evaluated using the given $\phi_i\big(\vx_i(T)\big)$ and terminal states $\vx(T)$ obtained by forward propagation of \eqref{eqn:fsde}. This is then used to train the deep LSTM network using optimizers such as Adam \cite{kingma2014adam}. Thus, deep FBSDEs is a self-supervised learning framework. Over iterations, as the loss is minimized, the network improves its predictions of $\hat{V}\big(\vx(0),\,0\big)$ and $\frac{\partial V}{\partial \vx}(\vx,\,t;\,\theta)$.
\par For unconstrained problems, $\vu_i^*=-\vR_i^{-1}\vG_i\T\frac{\partial V}{\partial \vx_i}$. However, when combined with SCBFs, $\vu_i^*$ can only be computed numerically. Similar to work in \cite{pereira2020safe}, safe optimal controls $\vu_i^*$ can be computed in an end-to-end differentiable manner using an OptNet-like \cite{amos2017optnet} implicit layer to solve  the constrained QP \eqref{eqn:constrained_centralized_HJB} and to ensure efficient backpropagation for DNN training using the implicit function theorem. 

\section{Decentralized Approach}
\label{sec:dec}
%
% \par\textcolor{red}{Mention somewhere in this section that, \textit{We first aim to decouple the individual Hamiltonian minimization problems ... . Also, mention reduction to $N(|\mathcal{N}|-1)$ (or $\sum_i^N (|\calN_i| - 1)$ since we haven't introduced the assumption yet) constraints.}}
In this section, we propose a decentralized approach for addressing the Hamiltonian minimization in \eqref{eqn:constrained_centralized_HJB}. To achieve this, we first reformulate the centralized problem in a distributed form. Subsequently, we propose a decentralized ADMM-based method combining elements from Consensus ADMM \cite{boyd2011distributed} and OSQP \cite{osqp} for solving large-scale QPs, which is then employed for solving our problem.

\subsection{Decentralized Problem}
%
% \textcolor{red}{(ensure that both $\mathcal{P}_i$ and $\mathcal{N}_i$ are defined such that they include agent $i$)}\\
The key restriction in problem \eqref{eqn:constrained_centralized_HJB} which prevents us from directly solving it in a distributed manner is the coupling induced by the inter-agent constraints. To overcome this issue, let us introduce the \textit{neighborhood} sets $\calN_i, \ i \in \llbracket 1, N \rrbracket$, which contain the indices of the neighboring agents of each agent $i$. For instance, in a 5-agent scenario where agents $2,4,5$ are neighbors of agent 1, its neighborhood set would be $\calN_1 = \{2,4,5\}$. We also define the sets $\calP_i = \{ j: i \in \calN_j \}, \ i \in \llbracket 1, N \rrbracket$, where each set $\calP_i$ contains all the agents that have agent $i$ as a neighbor. 

\begin{assumption}
\label{assumption1}
Each agent $i \in \llbracket 1, N \rrbracket$ is able to communicate with all agents $j \in \calN_i \cup \calP_i$, and vice versa.
\end{assumption}

Next, we consider for each agent $i$, the copy control and state variables $\{ \vu_j^{(i)} \}_{j \in \calN_i}$ and $\{ \vx_j^{(i)} \}_{j \in \calN_i}$, respectively. Essentially, a copy variable $\vu_j^{(i)}$ can be interpreted as \textit{agent $i$ deciding what is safe for its neighbor $j$ from its own perspective}. Let us also define the augmented local variables containing the states and controls of all agents within agent $i$'s neighborhood:
\begin{equation}
\tilde{\vu}_i = 
\big[
\vu_i; \{ \vu_j^{(i)} \}_{j \in \calN_i}
\big], \ 
\tilde{\vx}_i = 
\big[
\vx_i; \{ \vx_j^{(i)} \}_{j \in \calN_i}
\big], \ i \in \llbracket 1, N \rrbracket.
\nonumber
\end{equation}
Nevertheless, the inclusion of the copy variables creates a requirement for enforcing a consensus between variables that correspond to the same agents. For this reason, we also introduce the global control variable $\vg = [\vg_1; \dots; \vg_N]$ and impose the following constraints between local and global variable components,
\begin{equation}
\vu_j^{(i)} = \vg_j, 
\ \forall j \in \calN_i \cup \{ i \}, \ \forall i \in \llbracket 1, N \rrbracket.
\end{equation}

We can now formulate a \textit{decentralized} form of the Hamiltonian minimization problem as,
\begin{subequations}
\label{Dec problem 1}
\begin{align}
& \quad \quad \quad \quad \ \min \sum_{i=1}^N \calH_i(\tilde{\vu}_i)
\label{Dec problem 1 - Cost}
\\ 
\text{s.t.} \quad  
& \dfrac{\partial B^{i, k}}{\partial \tilde{\vx}_{i,k}} 
\big(\tilde{\vf}_{i,k} + \tilde{\vG}_{i,k} \tilde{\vu}_{i,k} \big) + \dfrac{1}{2} \tr \bigg(\dfrac{\partial^2 B^{i, k}}{\partial \tilde{\vx}_{i,k}^2} \tilde{\vSigma}_{i,k} \tilde{\vSigma}_{i,k} \T \bigg) 
\nonumber
\\
& \leq 
-\alpha B^{i,k} + \beta
, \ \forall k \in \llbracket 1, N_{\text{ineq},i} \rrbracket, \ \forall i \in \llbracket 1, N \rrbracket
\label{Dec problem 1 - Constraints}
\\ 
& \tilde{\vu}_i = \tilde{\vg}_i, \ \forall i \in \llbracket 1, N \rrbracket
\label{Dec problem 1 - Consensus constraints}
\end{align}
\end{subequations}
where the vectors $\tilde{\vx}_{i,k}$, $\tilde{\vu}_{i,k}$ are defined as 
$\tilde{\vx}_{i,k} = [\vx_i; \vx_j^{(i)}]$, $\tilde{\vu}_{i,k} = [\vu_i; \vu_j^{(i)}]$, 
if \eqref{Dec problem 1 - Constraints} is an inter-agent constraint involving a specific neighboring agent $j \in \calN_i$ and as 
$\tilde{\vx}_{i,k} = \vx_i$, $\tilde{\mathbf{u}}_{i,k} = \mathbf{u}_i$,
if \eqref{Dec problem 1 - Constraints} is an obstacle avoidance constraint. The functions $\tilde{\vf}_{i,k}$, $\tilde{\vG}_{i,k}$, $\tilde{\vSigma}_{i,k}$ are defined accordingly in each case. 
Finally, $\tilde{\vg}_i$ is defined as 
$\tilde{\vg}_i = 
\big[
\vg_i; \{ \vg_j \}_{j \in \calN_i}
\big] 
$ and $N_{\text{ineq},i} = |\calN_i| + N_o$.
% $(\tilde{\vg}_i)_j = \vg_{M(i,j)}$. 

Subsequently, by denoting the linear inequality constraints \eqref{Dec problem 1 - Constraints} as $\vA_i \tilde{\vu}_i \leq \vd_i$, we can rewrite problem \eqref{Dec problem 1} in a more compact form as,
\begin{align}
\min & \sum_{i=1}^N \calH_i(\tilde{\vu}_i) + \calI_{\vA_i \tilde{\vu}_i \leq \vd_i}(\vA_i \tilde{\vu}_i)
% \label{Dec problem 2 - Cost}
\nonumber
\\ 
\text{s.t.} \quad & \tilde{\vu}_i = \tilde{\vg}_i, \ \forall i \in \llbracket 1, N \rrbracket.
\label{Dec problem 2}
\end{align}
Problem \eqref{Dec problem 2} is now in a form where CADMM could be directly applied to solve it. This would yield a bilevel distributed optimization algorithm where at every ADMM iteration, each agent would first locally solve a QP, and then the local solutions would be used to perform the global and dual updates \cite[Chapter 7]{boyd2011distributed}. These local QPs could be solved by well-known solvers such as OSQP \cite{stellato2020osqp}, interior-point methods \cite{nocedal2006numerical, mattingley2012cvxgen,amos2017optnet}, etc. 
% \oswin{Maybe write down the CADMM iterations here so that you can compare with the merged OSQP iterations below?}
%
\subsection{Merged CADMM-OSQP Method}
% \augustinos{TODO: Change dual varialb to $\tau$, $\tau$ to $k$ (discretization) and pen. parameters to $\rho_1$ and $\rho_2$.}
In this work, we exploit the fact that the inner QP program in CADMM could itself be solved using the ADMM-based solver OSQP. Therefore, we propose 
flattening the bilevel optimization framework with a novel Merged CADMM-OSQP method for solving QPs in a decentralized manner. 

First, let us define $\tilde{\vR}_i = \blkdiag(\vR_i, \{ \mathbf{0} \}_{k \in \llbracket 1, |\calN_i| \rrbracket })$ and $\tilde{\vq}_i = [ \vq_i; \{ \mathbf{0} \}_{k \in \llbracket 1, |\calN_i| \rrbracket }]$ where $\vq_i=\frac{\partial V}{\partial \vx_i}\T\vG_i$. We can then reformulate \eqref{Dec problem 2} as,
\begin{align}
& \min \sum_{i=1}^N \frac{1}{2} \tilde{\vu}_i \T \tilde{\vR}_i \tilde{\vu}_i + \tilde{\vq}_i \T \tilde{\vu}_i + \calI_{\tilde{\vz}_i \leq \vd_i}(\tilde{\vz}_i)
\nonumber
\\ 
& ~ \text{s.t.} \quad 
\vA_i \tilde{\vu}_i = \tilde{\vz}_i, \ \tilde{\vu}_i = \tilde{\vg}_i, \  \forall i \in \llbracket 1, N \rrbracket 
\label{Dec problem 3}
\end{align}
where each $\calH_i(\tilde{\vu}_i)$ is given by $\calH_i(\tilde{\vu}_i) = \frac{1}{2} \tilde{\vu}_i \T \tilde{\vR}_i \tilde{\vu}_i + \tilde{\vq}_i \T \tilde{\vu}_i$. Next, we introduce the auxiliary variables $\hat{\vz}_i, \ i \in \llbracket 1, N \rrbracket$ in a similar manner as in the OSQP derivation \cite[Section 3]{stellato2020osqp} and transform \eqref{Dec problem 3} to,
\begin{align}
\min & \sum_{i=1}^N \frac{1}{2} \tilde{\vu}_i \T \tilde{\vR}_i \tilde{\vu}_i + \tilde{\vq}_i \T \tilde{\vu}_i 
+ \calI_{\vA_i \tilde{\vu}_i = \tilde{\vz}_i}(\tilde{\vu}_i, \tilde{\vz}_i)
+ \calI_{\tilde{\vz}_i \leq \vd_i}(\hat{\vz}_i)
\nonumber
\\ 
& ~~~~~~~~ \text{s.t.} \quad 
\tilde{\vz}_i = \hat{\vz}_i, \ \tilde{\vu}_i = \tilde{\vg}_i, \ \forall i \in \llbracket 1, N \rrbracket.
\label{Dec problem 4}
\end{align}
%
% \begin{subequations}
% \label{Dec problem 4}
% \begin{align}
% \min & \sum_{i=1}^N \calH_i(\tilde{\vu}_i) 
% + \calI_{\vA_i \tilde{\vu}_i = \tilde{\vz}_i}(\tilde{\vu}_i, \tilde{\vz}_i)
% + \calI_{\calC_i}(\hat{\vz}_i)
% \label{Dec problem 4 - Cost}
% \\ 
% \text{s.t.} \quad 
% & (\tilde{\vu}_i, \tilde{\vz}_i) = (\hat{\vu}_i, \hat{\vz}_i), \ \tilde{\vu}_i = \tilde{\vg}_i, \ \forall i \in \llbracket 1, N \rrbracket
% \label{Dec problem 4 - Consensus constraints}
% \end{align}
% \end{subequations}
%
The Augmented Lagrangian (AL) of \eqref{Dec problem 4} yields,
\begin{align}
\calL = 
& \sum_{i=1}^N \frac{1}{2} \tilde{\vu}_i \T \tilde{\vR}_i \tilde{\vu}_i + \tilde{\vq}_i \T \tilde{\vu}_i
+ \calI_{\vA_i \tilde{\vu}_i = \tilde{\vz}_i}(\tilde{\vu}_i, \tilde{\vz}_i)
+ \calI_{\tilde{\vz}_i \leq \vd_i}(\hat{\vz}_i)
\nonumber
\\
& + \frac{\rho_1}{2} \Big\| \tilde{\vz}_i - \hat{\vz}_i + \frac{\vy_i}{\rho_1} \Big\|_2^2
+ \frac{\rho_2}{2} \Big\| \tilde{\vu}_i - \tilde{\vg}_i + \frac{\bzeta_i}{\rho_2} \Big\|_2^2
\end{align}
%
% %
% \begin{align}
% \calL = 
% & \sum_{i=1}^N \calH_i(\tilde{\vu}_i) 
% + \calI_{\vA_i \tilde{\vu}_i = \tilde{\vz}_i}(\tilde{\vu}_i, \tilde{\vz}_i)
% + \calI_{\calC_i}(\hat{\vz}_i)
% \nonumber
% \\
% & + \frac{\rho}{2} \Big\| \tilde{\vu}_i - \hat{\vu}_i + \frac{\vy_i}{\rho} \Big\|_2^2
% + \frac{\sigma}{2} \Big\| \tilde{\vz}_i - \hat{\vz}_i + \frac{\blambda_i}{\sigma} \Big\|_2^2
% \nonumber
% \\ 
% & + \frac{\mu}{2} \Big\| \tilde{\vu}_i - \tilde{\vg}_i + \frac{\bxi_i}{\mu} \Big\|_2^2
% \end{align}
% %
where $\vy_i$, $\bzeta_i$ are the dual variables for the corresponding equality constraints and $\rho_1, \rho_2 >0$ are penalty parameters. 
% Note that in contrast to \cite{stellato2020osqp}, there is no need to introduce an additional constraint $\tilde{\vu}_i = \hat{\vu}_i$ (and an equivalent penalty parameter let us say $\sigma$) since the facts that the matrices $\vR_i$ are positive definite and $\mu >0$ guarantee that the KKT system for the ADMM subproblem \eqref{Merged Block 1} will be well-defined. \oswin{What is $\hat{\vu}$?}

Therefore, it is possible to use ADMM in a manner such that the consensus and OSQP updates take place within the same cycle of updates. In the following, the superscript $l \in \llbracket 1, l_{\text{max}} \rrbracket$ denotes the current ADMM iteration, with $l_{\text{max}}$ being the maximum allowed iterations. The first block of updates on $\tilde{\vu}_i$ and $\tilde{\vz}_i$ will be,
\begin{align}
\{ \tilde{\vu}_i, \tilde{\vz}_i \}^{l+1} = 
& \argmin_{ \tilde{\vu}_i, \tilde{\vz}_i }
\frac{1}{2} \tilde{\vu}_i \T \tilde{\vR}_i \tilde{\vu}_i + \tilde{\vq}_i \T \tilde{\vu}_i
\nonumber
\\
& 
+ \frac{\rho_1}{2} \Big\| \tilde{\vz}_i - \hat{\vz}_i^l + \frac{\vy_i^l}{\rho_1} \Big\|_2^2
+ \frac{\rho_2}{2} \Big\| \tilde{\vu}_i - \tilde{\vg}_i^l + \frac{\bzeta_i^l}{\rho_2} \Big\|_2^2
\nonumber
\\[0.2cm]
& \text{s.t. } \vA_i \tilde{\vu}_i = \tilde{\vz}_i.
\label{Merged Block 1}
\end{align}
The second block where $\hat{\vz}_i$ and $\vg$ are updated will consist of,
\begin{subequations}
\label{Merged Block 2}
\begin{align}
% \hat{\vz}_i^{l+1} & = \kappa \tilde{\vz}_i^{l+1} + (1 - \kappa) \hat{\vz}_i^l + \frac{1}{\sigma} \blambda_i^l 
% \\
% \hat{\vz}_i^{l+1} & = \Pi_{\calC_i}( \kappa \tilde{\vz}_i^{l+1} + (1 - \kappa) \hat{\vz}_i^l + \frac{1}{\rho} \vy_i^l ) 
\hat{\vz}_i^{l+1} & = \Pi_{(-\infty, \vd_i]}( \tilde{\vz}_i^{l+1} + \frac{1}{\rho_1} \vy_i^l ) 
\\
% \vg_m^{l+1} & = \frac{1}{\kappa_m} \sum_{M(i, j) = m} \left( (\tilde{\vu}_i^{l+1})_j + \frac{1}{\mu} (\bxi_i^l)_j \right)
\vg_i^{l+1} & = \frac{1}{|\calP_i| + 1} \sum_{j \in \calP_i \cup \{ i \}} \left( \vu_i^{(j),l+1} + \frac{1}{\rho_2} \bzeta_i^{(j),l } \right)
\label{Merged Block 2 b}
\end{align}
\end{subequations}
where $\bzeta_i^{(j)}$ is the part of the dual variable $\bzeta_i$ corresponding to the constraint $\vu_j^{(i)} = \vg_j$. 
Finally, the dual variables updates will be,
\begin{subequations}
\label{Merged Block 3}
\begin{align}
% \blambda_i^{l+1} & = \blambda_i^l + \sigma ( \kappa \tilde{\vz}_i^{l+1} + (1 - \kappa) \hat{\vz}_i^l - \hat{\vz}_i^{l+1})
% \\
\vy_i^{l+1} & = \vy_i^l + \rho_1 ( \tilde{\vz}_i^{l+1} - \hat{\vz}_i^{l+1} ) 
\\
\bzeta_i^{l+1} & = \bzeta_i^{l} + \rho_2 ( \tilde{\vu}_i^{l+1} - \tilde{\vg}_i^{l+1} ).
\label{Merged Block 3 b}
\end{align}
\end{subequations}

Since updating each variable with subscript $i$ only requires variables that have the same subscript, the updates \eqref{Merged Block 1}, \eqref{Merged Block 2}, \eqref{Merged Block 3} can be performed in parallel by each agent. Therefore, the algorithm can be executed in a fully decentralized manner. During every ADMM iteration, two communication steps are required. The first one takes place before the computation of \eqref{Merged Block 2 b}, where every agent $ j \in \calP_i$ must send its variables $\tilde{\vu}_j^{l+1}$ and $\bzeta_j^l$ to each agent $i$. The second one is performed before \eqref{Merged Block 3 b}, where every agent $ j \in \calN_i$ must send $\vg_j^{l+1}$ to agent $i$, so that the latter can construct the vector $\tilde{\vg}_i^{l+1}$. Finally, one extra communication step is required before the first ADMM iteration between agent $i$ and its neighbors $j\in \mathcal{N}_i$, to construct $\vA_i$ and $\vd_i$ for the inequality constraints which remain the same for all subsequent ADMM iterations.
\par It should be highlighted that the $N_{\text{ineq},i} = r + N_o$ constraints involved in each of the local subproblems, will be drastically fewer than the $N_{\text{ineq}} = {N \choose 2} + N N_o$ constraints of the centralized problem for $r \ll N$. 
% \oswin{Maybe an algorithm + figure to describe the computations done + flow of information would be useful?}

%

% {\color{blue}
% \begin{enumerate}
% \item 1st subsection. Define copy (and augmented) variables
% \item Rewrite the Hamiltonian minimization containing the copy variables
% \item Introduce global variables to enforce consensus
% \item 2nd subsection. Mention that is problem is a QP. Add one OSQP copy variable. Explain why we don't need x. 
% \item Write AL.
% \item Derive ADMM update equations.
% \end{enumerate}

% }
\section{Implementation Details}
We make the following assumption for our implementation,
\begin{assumption}
\label{ass:fixed_neighborhood}
All neighborhood sets within a single time step $t$,  $\calN_i(t)$, are of equal size $r$.
\end{assumption}
This assumption ensures that all $\Nineqi$ are equal so that each local subproblem can be solved in a batched fashion,
thereby allowing for efficient training on GPUs.
Note that this is not a very restrictive assumption as we still allow the individual $\calN_i(t)$ to change across time.

%
% \begin{remark}
% \label{assumption2}
% { \color{gray}
% The neighbors of agent $i$, is determined by its $\big(|\mathcal{N}_i|-1\big)$ closest neighbors.
% }
% \end{remark}
%
% \begin{remark}
% \label{assumption3}
% { \color{gray}
% The neighborhood size of each agent $i$ is fixed (i.e, $|\mathcal{N}_i|(t)=$constant) for the entire time horizon $T$.
% }
% \end{remark}
\begin{figure}[t]
    \centering
    \resizebox{\columnwidth}{!}{%
    \begin{tikzpicture}[mybackground={here},scale=0.4]{sans text}
    \node (rndcp) at (0, 0) [fill=mat_blue!30,minimum width=3cm,minimum height=1.5cm,align=center]
    {Reduced Non-Duplicate\\Centralized Problem\\(RNDCP)};
    
    \node (dp) at (16, 0) [fill=mat_red!30,minimum width=3cm,minimum height=1.5cm,align=center]
    {Decentralized Problem};
    
    \node (rdcp) at (16, -7) [fill=mat_green!30,minimum width=3cm,minimum height=1.5cm,align=center]
    {Reduced Duplicate\\Centralized Problem\\(RDCP)};
    
    \node [above=0mm of rndcp] {\textbf{Original Problem}};
    \node [above=0mm of dp] {\textbf{Forward Pass}};
    \node [below=0mm of rdcp] {\textbf{Backward Pass}};
    
    \draw [->, line width=0.5mm] (rndcp) to node[above,align=center]{Cheaper forward\\pass} (dp);
    \draw [->, line width=0.5mm] (dp) to node[right,align=center]{Cheaper backward pass\\(No copy variables)} (rdcp);
    \draw [->, line width=0.5mm] (rndcp) to node[left,align=center,xshift=0mm,yshift=-3mm]{Cheaper backward pass\\(Don't need to form unique constraints)} (rdcp);
    \end{tikzpicture}
    }
    \vspace*{-5mm}
    \caption{Relationship between RNDCP,
    RDCP \eqref{eqn:reduced_duplicate_centralized_problem} and the Decentralized Problem (Section \ref{sec:dec}).}
    \label{fig:relationship_between_problems}
\end{figure}
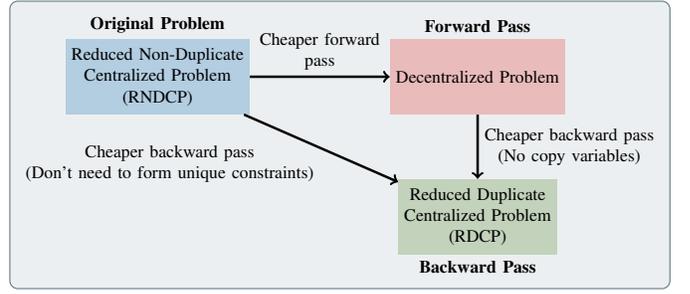

\subsection{Time Discretization}
In order to use deep learning, we consider a Euler-Maruyama time discretization of \eqref{eqn:fsde} and \eqref{eqn:bsde} so that backpropagation through time can be performed on a finite number of time steps to train the deep network. This is similar to past deep FBSDE works \cite{pereira2019learning, pereira2020feynman, pereira2020safe} wherein using a finite time interval of $\Delta t$, the time horizon is divided into $\frac{T}{\Delta t}$ equal intervals of length $\Delta t$. The time-discretized equations are, 
\begin{align}
    \vx[\tau + 1] &= \vx[\tau] + (\vf + \vG\vu^*)\Delta t + \vSigma \sqrt{\Delta t}\,\bepsilon\label{eqn:discretized_FSDE}\\
    V[\tau + 1] &= V[\tau]  -\bigg[\sum_{i=1}^N\Big(c_i+\frac{1}{2}\vu_i^* {}\T\vR_i\vu_i^*\Big)\bigg]\Delta t \nonumber\\&\qquad\qquad+ \frac{\partial V}{\partial \vx}\T \vSigma \sqrt{\Delta t}\,\bepsilon 
    \label{eqn:discretized_BSDE}
\end{align}
where $\bepsilon \sim\mathcal{N}(\mathbf{0}, \vI)$ and $\tau \in \llbracket 0,\frac{T}{\Delta t} -1 \rrbracket$.
\subsection{Forward Pass}
\begin{figure*}[ht!]
    \centering
    \includegraphics[width=0.92\linewidth]{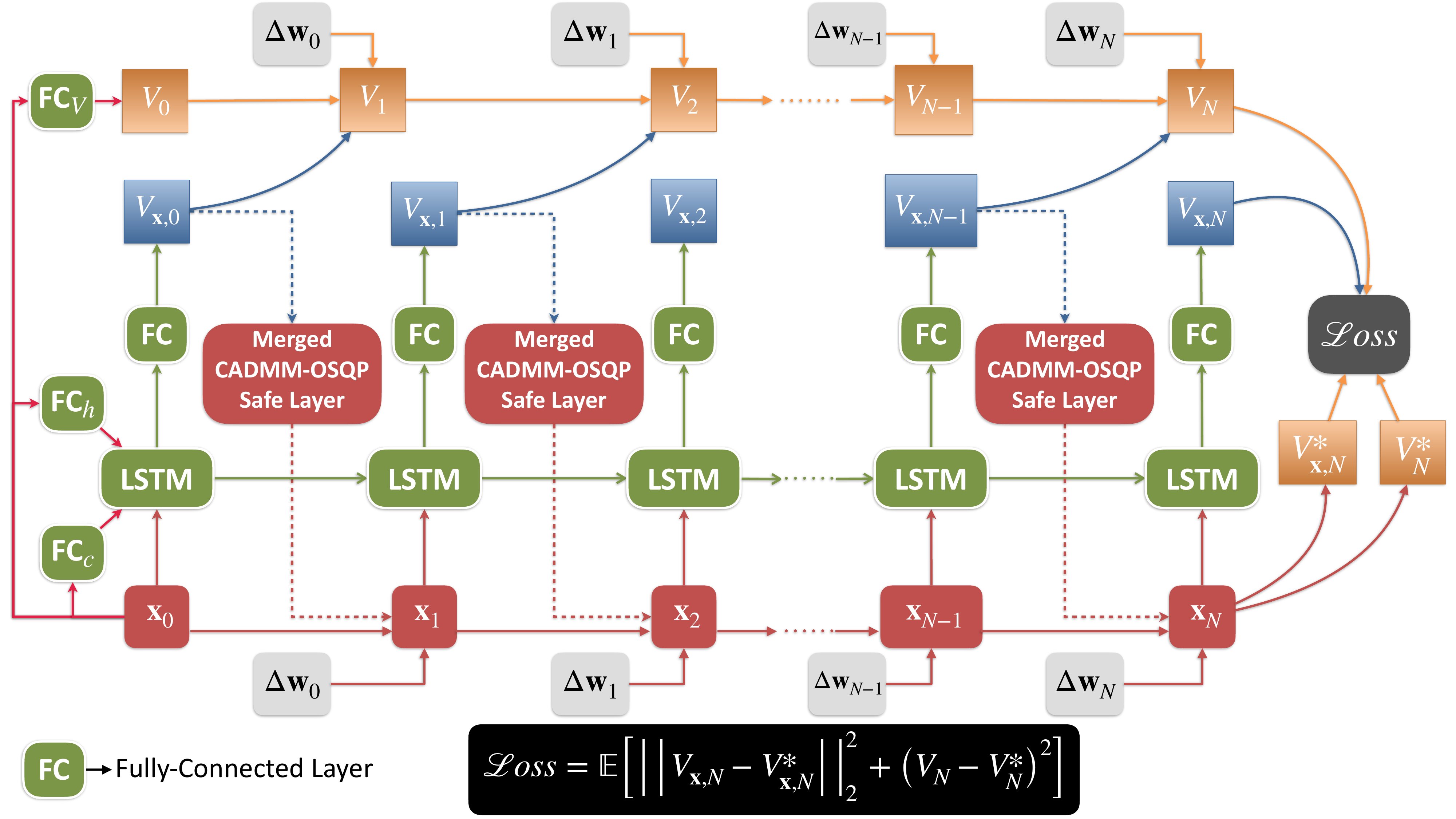}
    \caption{ Unrolled Deep FBSDE compute graph using the proposed Merged CADMM-OSQP implicit safe layer. }
    \label{fig:network_arch}
\end{figure*}

Similar to past work \cite{pereira2019learning, pereira2020safe}, the forward pass involves propagating the discretized FBSDE and BSDE forward in time using \eqref{eqn:discretized_FSDE} and \eqref{eqn:discretized_BSDE}. The primary distinction between \cite{pereira2020safe} and our approach is a new implicit safe layer based on CADMM. This difference is clear by comparing the unrolled compute graph shown in Figure \ref{fig:network_arch} with that of \cite[Figure 1]{pereira2020safe}. The additional difference from past works is the inclusion of extra fully-connected layers $\text{FC}_c, \text{FC}_h$ and $\text{FC}_V$ to allow for training from random initial conditions. These extra networks serve to initialize the initial cell-state and initial hidden-state of the LSTM layers and the initial value-function respectively. 

\subsection{Backward Pass}
Our proposed decentralized Merged CADMM-OSQP solver is an instantiation of an implicit neural network layer. Hence, we utilize the implicit function theorem to compute the necessary gradients for the backward pass. Now, one approach could be to formulate the KKT matrix resulting from the KKT conditions that the optimal solution, $(\tilde{\vu}_i^*,\,\vy_i^*)\,\,\forall i\in \llbracket1, N\rrbracket$, must satisfy. This is precisely the solution of the decentralized problem in the forward pass. However, the resulting size of the KKT matrix could be very large depending on $N$ and $r$ because of the presence of control copy variables for each agent's neighbors used in the decentralized problem. Based on the insight that each $\tilde{\vu}_i^*$ also satisfies the consensus constraint, $\tilde{\vu}_i^*=\tilde{\vg}_i$, we can eliminate the copy variables and  consider only the control variables of the ego agents of each neighborhood.  However, if two agents $i,j$ are mutual neighbors, i.e., $j \in \calN_i$ and $i \in \calN_j$, then the constraint involving $i$ and $j$ will appear twice in the constraint matrix (i.e., as duplicate row entries). We therefore refer to this problem as the \textit{Reduced Duplicate Centralized Problem} (RDCP) where \textit{reduced} indicates the reduction in the number of constraints from $N_{\text{ineq}}$ to $N N_{\text{ineq},i}$. This is formally stated as follows: \begin{equation}
 \min_\vu \,\,\mathcal{H}(\vu),\quad\text{s.t. } \vC \vu \leq \vd
\label{eqn:reduced_duplicate_centralized_problem}   
\end{equation}
where $\vd=[\vd_1;\,\vd_2;\,\ldots;\,\vd_N]$ and $\vC$ can be constructed from the neighborhood constraint matrices $\vA_i$ (see supplementary material \Cref{sec:construction_C_matrix}). Using \Cref{eqn:reduced_duplicate_centralized_problem} would result in a much smaller KKT matrix and thus a smaller system of equations to solve for the backward pass gradients. This is what is implied by \textit{cheaper backward pass} on the arrow going from the decentralized problem to the RDCP in \Cref{fig:relationship_between_problems}. 
\par In contrast to the RDCP is the \textit{Reduced Non-Duplicate Centralized Problem} (RNDCP) wherein the objective stays the same as in \Cref{eqn:reduced_duplicate_centralized_problem}, however, the constraint matrix is now $\bar{\vC}$ which has only the unique entries of the constraints forming $\vC$. There are two reasons we choose the RDCP over the RNDCP for solving the backward pass:
\begin{enumerate}
    \item Construction of the non-duplicate constraint matrix $\bar{\vC}$ requires checking if mutual neighbors exist for every agent which does not scale for large $N$,
    \item $\vC$ can be easily constructed using the matrices $\vA_i$ used in the forward pass of Merged CADMM-OSQP.
\end{enumerate}
To justify the usage of \eqref{eqn:reduced_duplicate_centralized_problem}, we first make the following assumption about the corresponding RNDCP:
% \begin{assumption}
% \label{assumption:rndcp_assumption}
% The QP of the non-duplicate problem satisfies the LICQ (Linear Independence Constraint Qualification), i.e.,
% the active constraints of the matrix $\bar{\vC}$ are linearly independent.
% \end{assumption}
\begin{assumption}
\label{assumption:lemma_assumption}
The control cost matrix $\vR$ is positive definite and LICQ holds for the reduced non-duplicate problem,  i.e.,
the active constraints of the matrix $\bar{\vC}$ are linearly independent.
\end{assumption}
Constraint qualifications are necessary for optimal solutions to constrained optimization problems to satisfy the KKT conditions.
LICQ is one of the most frequently used constraint qualification in optimization literature \cite{nocedal2006numerical}. In our case, we additionally rely on LICQ to ensure that the Lagrange multipliers satisfying the KKT conditions are unique \cite[Section 3]{wachsmuth2013licq}.
Given that \Cref{assumption:lemma_assumption} holds, we can establish the connection between the unique Lagrange multipliers of the RNDCP and those of the RDCP by the following lemma,
%
% \begin{customlemma}{1}
% Let $\mathcal{D} \coloneqq \{ \mathcal{D}_i \}_{i=1}^\Nineqb$
% with $\mathcal{D}_i \coloneqq \{ \vcdij \}_{j=1}^{\xi_i}$
% denote the set of equivalence classes induced by the equality equivalence relation
% \begin{equation}
%     \vc_{\zeta_{i,a}} \sim \vc_{\zeta_{i,b}} \iff
%     \vc_{\zeta_{i,a}} = \vc_{\zeta_{i,b}}, \quad \forall a, b \in \llbracket 1, \xi_i \rrbracket \;
%         \forall i
% \end{equation}
% where $\zeta_{i, j}$ denotes the row corresponding to the $j$-th instance of the $i$-th unique constraint.
% In other words, there are $\xi_i$ duplicates for each of the $\Nineqb$ unique constraint rows.
% Suppose that $\vR$ is positive definite and LICQ holds for the non-duplicate problem.
% Then,
% \begin{align}
%     \lambdanondup{i} &= \sum_{j=1}^{\xi_i} \lambdadup{\zeta_{i,j}}
%     \label{eq:relation_dup:sum_lambda} \\
%     \lamdlamnondup{i} &= \sum_{j=1}^{\xi_i} \lamdlamdup{\zeta_{i,j}}
%     \label{eq:relation_dup:sum_lambdadlambda}
% \end{align}
% where $\blambdadup, \blambdanondup$ denote the Lagrange multipliers and $\dblambdadup, \dblambdanondup$ denote the variables of the KKT system for the duplicate and non-duplicate problems respectively.
% \end{customlemma}
%

% The above lemma then allows to prove the following theorem.
\begin{customlemma}{1}
\label{lemma:lemma_1}
Given that Assumption \ref{assumption:lemma_assumption} holds, then it follows that,
\begin{align}
    \lambdanondup{l} &= \sum_{j=1}^{\xi_l} \lambdadup{\zeta_{l,j}} \quad \forall l \in \llbracket 1, \Nineqb  \rrbracket
    \label{eq:relation_dup:sum_lambda} 
\end{align}
where $\blambdadup, \blambdanondup$ denote the vectors of Lagrange multipliers 
for the duplicate and non-duplicate problems respectively.
\end{customlemma}
In the following lemma, we derive the KKT system for the RDCP as well as expressions for the gradients of the loss function $\ell$ with respect to parameters of interest of the Merged CADMM-OSQP Safe Layer for any given every time step $t$.
\begin{customlemma}{2}[QP Gradients]
\label{lemma:lemma_2}
Let $\vu^*$ and $\blambdadup^*$ denote the solutions to the reduced duplicate problem at any given time step $t$, 
\begin{align}
    \min_{\vu(t)} &\quad \frac{1}{2} \vu(t)\T \vR(t) \vu(t) + \vq(t)\T \vu(t) \\
    \textrm{s.t.} &\quad \vC(t) \vu(t) \leq \vd(t)
\end{align}
and let $\ell$ denote the overall neural-network training loss function.
Then, the gradients of $\ell$ with respect to the data matrices $\vR, \vq, \vC, \vd$ for the time step $t$ have the form,
\begin{subequations}
\begin{align}
\nabla_\vq \ell  &= \dvu
\label{eq:grad_q} \\
\nabla_{\vd} \ell  &= -\diag(\blambdadup^*) \dblambdadup
\label{eq:grad_d} \\
\nabla_\vR \ell &= \frac{1}{2}(\dvu \, {\vu^*}\T + \vu^* \, \dvu\T)
\label{eq:grad_R} \\
\nabla_{\vC} \ell &= \blambdadup^* \, \dvu\T + \diag(\blambdadup^*) \dblambdanondup {\vu^*}\T 
\label{eq:grad_C}
\end{align}
\end{subequations}
where $\dvu$ and $\dblambdanondup$ are the solutions to the KKT system
\begin{equation}
    \begin{bmatrix}
    \vR & \vC\T \diag(\blambdadup^*) \\
    \vC & \diag(\vC \vu^* - \vd)
    \end{bmatrix}
    \begin{bmatrix}
    \dvu \\ \dblambdadup
    \end{bmatrix}
    =
    \begin{bmatrix} -\nabla_{\vu^*} \ell \\ \mathbf{0} \end{bmatrix}.
\end{equation}
\end{customlemma}
Having derived the KKT system for RDCP in \Cref{lemma:lemma_2}, we can establish a relation between the KKT variables of the RNDCP and those of the RDCP in the following lemma,
\begin{customlemma}{3}
\label{lemma:lemma_3}
Given that Assumption \ref{assumption:lemma_assumption} holds, then it follows that,
\begin{align}
    \lamdlamnondup{l} &= \sum_{j=1}^{\xi_l} \lamdlamdup{\zeta_{l,j}},\, \forall l \in \llbracket 1, \Nineqb  \rrbracket
    \label{eq:relation_dup:sum_lambdadlambda}
\end{align}
where 
% $\blambdadup, \blambdanondup$ denote the Lagrange multipliers 
$\dblambdadup, \dblambdanondup$ denote the vectors of the variables of the KKT systems for the duplicate and non-duplicate problems respectively.
% \\\marcus{in the main paper we use different symbols for parameters e.g., d instead of zeta. We should adopt these in the main paper}
\end{customlemma}
We invite the interested reader to refer to \Cref{sec:lemmas_proofs_section} in the supplementary material for proofs of the lemmas stated above. We use these lemmas to relate the gradients of $\ell$ with respect to the parameters of interest of the RNDCP with those of the RDCP with the following theorem,
\begin{customthm}{1} \label{thm:same_gradient}
Let $\vM$ denote the matrix describing the relationship between the duplicate and non-duplicate constraints:
\begin{equation}
  \vC = \vM \bar{\vC}, \quad \vd = \vM \bar{\vd} \label{eq:M_def}
\end{equation}
Then, for loss $\ell$, the gradients $\nabla_\vR \ell, \nabla_\vq \ell, \nabla_\vC \ell, \nabla_\vd \ell$
coincide for the duplicate and non-duplicate problems and are unique.
\end{customthm}
\begin{proof}[Proof Sketch]
    We first show that the gradients
    $\nabla_\vR \ell, \nabla_\vq \ell, \nabla_{\bar{\vC}} \ell, \nabla_{\bar{\vd}} \ell$
    depend only on $\dvu$ and the sums of $ \lambdadup{\zeta_{l,j}}$ and $\lamdlamdup{\zeta_{l,j}}$ associated to each unique constraint $l$ of matrix $\bar{\vC}$.
    Applying Lemmas \ref{lemma:lemma_1} and \ref{lemma:lemma_3} then shows that the theorem holds.
\end{proof}
We refer the reader to \Cref{sec:theorem_proof_section} in the supplementary material for the complete proof. Finally, in \Cref{sec:relation_between_decentralized_and_RNDCP_section} we establish a connection between the optimal solutions of the decentralized problem and the RNDCP similar to that between the RDCP and the RNDCP. This allows us to conclude that the solutions of the decentralized problem must coincide with those of the RDCP and thereby justifies using the solution $(\tilde{\vu}^*, \vy^*)$ of the decentralized problem computed in the forward pass to formulate the KKT conditions of the RDCP for the computation of the backward pass gradients.
\par In practice, there may be situations at certain time steps when Assumption \ref{assumption:lemma_assumption} is violated,
in which case the computed gradient serves as a noisy version of the true gradient.

\subsection{Termination Criteria and Penalty Parameters Adaptation}
The algorithm terminates when the norms of the primal and dual residuals get below their corresponding tolerance levels,
\begin{equation*}
r_{pri,a} \leq \epsilon_{pri,a}, \quad r_{dual,a} \leq \epsilon_{dual,a}, \quad \forall a = 1,2.
\end{equation*}
where $a=1,2$ refer to the QP and consensus residuals, respectively. Detailed expressions for the residual norms $r_{pri,a}, \ r_{dual,a}$ and the tolerances $\epsilon_{pri,a}, \ \epsilon_{dual,a}$ are provided in Section \ref{sec: admm residuals} of the supplementary material.

The selection of the penalty parameters $\rho_1$ and $\rho_2$ is important since the former encourages the satisfaction of the local constraints of each agent subproblem, while the latter encourages achieving consensus. Low values of these parameters could result to slow convergence of the algorithm, thus requiring a large value of the ADMM maximum iterations parameter $l_{\text{max}}$. On the other hand, if their values are too high then their corresponding terms in \eqref{Merged Block 1} will dominate the objective function. From a practical standpoint, we accommodate for these issues by adopting the following adaptation schemes for the penalty parameters:
\begin{equation*}
\rho_1^{l+1} = \rho_1^l 
\sqrt{
\frac{ 
r_{pri,1}^l / \kappa_{pri,1}^l
}{
r_{dual,1}^l / \kappa_{dual,1}^l
}
}, \quad 
\rho_2^{l+1} = \rho_2^l 
\sqrt{
\frac{ 
r_{pri,2}^l / \kappa_{pri,2}^l 
}{
r_{dual,2}^l / \kappa_{dual,2}^l
}
}.
\end{equation*}
This scheme is inspired by the adaptation rules that are used in the OSQP solver \cite[Section 5.2]{stellato2020osqp}.
Note that the specific choice of termination criteria and adaptation rules employed in this work would require computation being performed by a central node. Future work aims at proposing termination criteria and adaptation rules that can be performed in a fully decentralized manner as well.

\subsection{Additional Constraints for Local Subproblems}

In the algorithm proposed in Section \ref{sec:dec}, achieving consensus---and thus ensuring the safety of the agents---fully relies on ADMM. To facilitate reaching a consensus, we suggest including in every subproblem of agent $i$: i) the obstacle avoidance constraints of its neighbors, and/or ii) the inter-agent constraints between its neighbors. The number of these additional constraints will be $r N_o$ and $r \choose 2$, respectively. Even after incorporating these constraints, we emphasize that, for $r \ll N$, the new $N_{\text{ineq},i} = r + N_o + r N_o + {r \choose 2}$ constraints in each agent's local QP will still be substantially smaller than the $N_{\text{ineq}} = {N \choose 2} + N N_o$ of the centralized problem. 

\section{Simulation Results}
\label{sec:simresults}
We test the proposed approach on a system consisting of multiple agents with unicycle dynamics (similar to \cite[Section V.B]{xie2017differential}) for any agent $i$ given by,
\begin{equation*}
    \dot{x}_i=v_i \cos(\theta_i),\quad \dot{y}_i=v_i \sin(\theta_i), \quad \dot{\theta_i}= v_i u_i^{\theta}, \quad \dot{v}_i= u_i^{v}.
\end{equation*}
Constructing a state vector $\vx_i=[x_i;\,y_i;\, \theta_i;\,v_i]$ and a control vector $\vu_i=[u_i^\theta;\,u_i^v]$ and assuming that noise only enters the acceleration channels, the stochastic dynamics for agent $i$ can be written as, 
\begin{equation*}
    \mathrm{d} \mathbf{x}_i = \vf_i\mathrm{d} t + \vG_i\mathbf{u}_i \mathrm{d} t + \vSigma_i \mathrm{d} \vw_i
\end{equation*}
where $\vf_i,\,\vG_i,\text{ and, }\vSigma_i$ are given by, \begin{equation*}
    \vf_i=\begin{bmatrix}
    v_i\cos(\theta_i)\\v_i\sin(\theta_i)\\0\\0
    \end{bmatrix},\, \vG_i=\begin{bmatrix}
    0 & 0\\0 & 0\\ v_i & 0 \\ 0 & 1
    \end{bmatrix},\,\vSigma_i = \begin{bmatrix} 0 & 0 & 0 & 0 \\ 0 & 0 & 0 & 0 \\ 0 & 0 & \sigma & 0 \\ 0 & 0 & 0 & \sigma \end{bmatrix}.
\end{equation*}
For our simulations we consider obstacle avoidance and collision avoidance safety constraints in four different types of tasks. These constraints are stated as functions of the positions of the agents and are therefore relative degree 2 (i.e., one needs to differentiate twice before the control shows up). However, the aforementioned SCBF in \eqref{eqn:safety_constraint} assumes that the relative degree of $h$ is 1. This is required to ensure that $\frac{\partial B}{\partial \vx}\T\vG$ is not zero. Therefore, the original position constraint, which we hereon refer to as $h_\text{pos}(\vx)$, must be modified to ensure that the modified function has relative degree 1. In \cite{pereira2020safe}, this modification takes the following form, \begin{equation}
    h(\vx) = h_\text{pos}(\vx) - \mu\,v^2.
    \label{eqn:old_h}
\end{equation}
The parameter $\mu$ controls how fast the system can safely move inside the safe set and thereby, the implication of adding the $-\mu\,v^2$ term introduces constraints on the velocity in addition to the original position constraint. Intuitively, when $h_\text{pos}=0$, for the system to stay safe (i.e., $h\geq0$), the only allowable safe velocity is $v=0$.
\par The main drawbacks of \eqref{eqn:old_h} are that it does not take into account the heading of the agents and that it penalizes positive and negative velocities equally. To overcome these we propose the following two types of barrier functions,
\begin{enumerate}
    \item \textbf{Type-A}: This is a pairwise safety constraint concerning two agents and is constructed as follows:\begin{align}
        &h^A(\vx) = h_\text{pos}^A(\vx) - \mu\,(v_i\,\IP_i + v_j\,\IP_j)  \\
        &\text{where, }h_\text{pos}^A(\vx) = \frac{1}{2}\Big((x_i - x_j)^2 +(y_i - y_j)^2 - 4r^2 \Big),\nonumber\\
        &\IP_i = \bar{p}_{ij}\T \bar{\theta}_i, \text{ and, } \IP_j = \bar{p}_{ji}\T \bar{\theta}_j\nonumber
    \end{align} 
    for any two agents $i$ and $j$, each with a radius of $r$. The inner-product ($\IP$) terms depend on vectors $\bar{p}_{ij}$ and $\bar{p}_{ji}$ which denote relative position vectors from $i$ to $j$ and $j$ to $i$ respectively and on the vectors $\bar{\theta}_i=[\cos\theta_i;\,\sin\theta_i]$ and $\bar{\theta}_j=[\cos\theta_j;\,\sin\theta_j]$ which denote unit vectors along the headings of agents $i$ and $j$ respectively. 
    \item \textbf{Type-B}: This type of safety constraint concerns an agent $i$ and an obstacle $o$. It is constructed as follows, \begin{align}
        &h^B(\vx) = h_\text{pos}^B(\vx) - \mu\,v_i\,\IP \\
        &\text{where, }\nonumber\\&h_\text{pos}^B(\vx) = \frac{1}{2}\Big((x_i - x_o)^2 +(y_i - y_o)^2 - (r_i+r_o)^2 \Big),\nonumber\\
        &\text{and }\IP = \bar{p}_{io}\T\theta_i
    \end{align}
    where, $(x_o,y_o)$ is the position of the obstacle's center and $r_o$ is the obstacle's radius. The vectors $\bar{p}_{io}$ and $\theta_i$ are defined similar to the type-A constraint above.
\end{enumerate}

Similar to work in \cite{pereira2020safe} our simulations are also conducted in a safe reinforcement learning setting where it is required to be safe not only during inference but also during the entire training process. The reader is encouraged to refer to the video\footnote{\url{https://youtu.be/qjPLUlaxJos}} to support this claim. The video shows the gradual emergence of optimal behavior as iterations progress, for each of the tasks described below, while staying safe during the entire training process. Additionally, it depicts the behavior of a single batch instance on the left frame and distributions over entire batches on the right frame wherein agents are depicted as particles. The frame on the right also demonstrates successful performance on average as justified by our choice of the mean (i.e., expected value) cost function \eqref{eqn:total_cost}. 

\subsection{Swapping Task}
We first consider the swapping cars task presented in \cite{pereira2020safe}. We show that our proposed approach is not only scalable to larger teams of cars but can handle the added complexity of obstacle avoidance. Each agent's goal is to swap positions with the diametrically opposite one while avoiding collisions. In Fig. \ref{fig:exp:one_obs_swap_16}, the distributions of the positions show that the agents avoid collisions with a high probability. On closer inspection, one can argue that the problem is highly symmetrical as no matter where you stand on the initial circle, each agent effectively solves a similar problem. To prove that our approach can handle more complex scenarios, we added extra obstacles to create an asymmetrical version of the same problem. The final policy is depicted in Fig. \ref{fig:exp:three_obs_swap_16}. As seen in the figure, the cars on the top right quadrant encounter obstacles much sooner along with encountering their neighbors. Observing the second plot of Fig. \ref{fig:exp:three_obs_swap_16}, we see that the final policy now leads to the agents circling around the new highly non-convex obstacle shape in order to complete the task. For all swapping tasks, we used a neighborhood size of $r=3$. For the symmetrical obstacle task, we used $r$ type-A and $1$ type-B constraints in every neighborhood. While for the unsymmetrical task, we used $r \choose 2$ type-A and $N_o(r+1)$ type-B constraints in every neighborhood. 

\subsection{Bottleneck Task}
For this task, the agents are required to pass through the bottleneck in the center (created by multiple circular obstacles) and achieve a desired formation on the other side of the bottleneck as seen in Fig. \ref{fig:exp:bottleneck_8}. To train this policy we designed the running cost $c_i(\vx)$, and terminal cost $\phi_i(\vx)$, such that the task is divided into two objectives - (i.) pass through the bottleneck, and, (ii.) achieve the desired formation. This was done to ``encourage'' the agents to prioritize passing through the bottleneck, before attempting to achieve the desired formation on the other side. Without the first objective, some agents may end up getting stuck in the highly non-convex regions between the circular obstacles. Thus, it helps to avoid these undesirable local minima. This objective was achieved by setting a target for the agents in the column close to the bottleneck at a point on the x-axis further away from the bottleneck and setting a target for the agents in the second column at a point on the x-axis close to the bottleneck. This \textit{intermediate target} was set for $80\%$ of the time horizon. For the remaining $20\%$, the targets were set so as to achieve the desired formation on the other side. For this task, we chose $r=3$ with each neighborhood containing $r$ type-A constraints between ego agent and neighbors and $6$ type-B constraints between the ego agent and the obstacles.

\subsection{Moving-obstacle (or uncooperative agent) Task}
The goal of this task is the same as the swapping task with the added complexity of a moving obstacle. The moving obstacle can be interpreted as an uncooperative agent (i.e., one that cannot be controlled) and moves from bottom to top along the y-axis as seen in Fig. \ref{fig:exp:moving_obs_8}. To train this policy, we first trained a policy containing the 8 agents without any obstacle on the swapping task. This pre-trained policy was then used as an initialization to train the policy to swap while avoiding an oncoming moving obstacle. The reason for this two step process being that a completely random initial policy is unable to safely explore in the presence of the moving obstacle. A common scenario that arises when one attempts to train the policy from scratch is that the agents try to directly move to their diametrically opposite targets but then come to a halt and congregate around the center because they encounter other agents. However, as the moving obstacle approaches, the closest agent to the obstacle is inevitably ``run-over" by the moving obstacle, as it has no where to escape. Using a pre-trained policy embeds the agents with the ability to ``move away from the moving obstacle's path" as they have pre-learned a behavior to initiate a coordinated turn in order to avoid colliding with other agents. For this task, we used a neighborhood size of $r=3$ with $r$ type-A constraints between the ego-agent and its neighbors and $r+1$ type-B constraints for obstacle avoidance.    

\subsection{Large-Scale Formation Task}
Here, we demonstrate the scalability of our framework by considering a task where a large-scale team of $32$ agents must achieve a desired rectangular formation. Each agent has $r=6$ neighbors and each local problem includes $r$ type-A constraints and $N_o(r+1) = 14$ type-B constraints. As shown in Fig. \ref{fig:exp:formation_64}, the distributions of the positions of agents successfully reach close to the desired targets.

\section{Discussion}

\subsection{Constraints Reduction and Increased Memory Efficiency}
In Table \ref{tab:constraints}, we compare the number of constraints considered by the centralized and decentralized approaches for each task presented in Section \ref{sec:simresults}. Clearly, there is a substantial reduction on the number of constraints when using the proposed approach,  implying that it is more scalable to large-scale systems than the equivalent centralized one. Table \ref{tab:mem} also demonstrates the reduced required memory usage and training iteration time of our approach.

\subsection{Low Position Constraint Violation} 
\begin{figure}
    \centering
    \includegraphics[width=\linewidth]{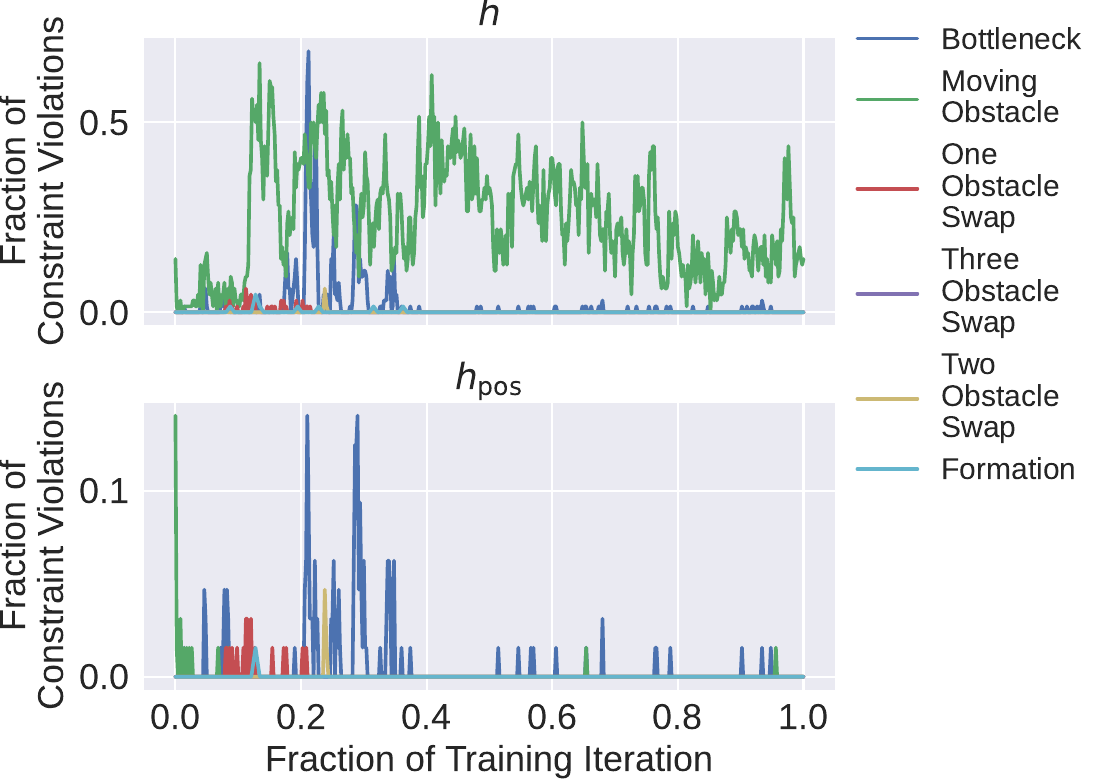}
    \caption{Constraint Violations}
    \label{fig:h_violations}
\end{figure}

In Figure \ref{fig:h_violations}, the top and bottom plots show the fraction of batch instances where $h<0$ and $h_\text{pos}<0$, respectively, against the number of training iterations. For all tasks, $h_\text{pos}<0$ occurs much less frequently as compared to $h<0$. This indicates that although there are many instances of $h$ violations (i.e., agents moving with \textit{unsafe} velocities in the vicinity of other agents or other obstacles), the number of physical collisions between agents or between agents and obstacles are much lower or even zero for some tasks. Thus, the new type-A and type-B barrier formulations allow for more aggressive behavior as compared to \eqref{eqn:old_h} which instead encourages conservative behaviors and therefore cannot scale for large numbers of agents. Another interesting observation is the sharp decline in $h_\text{pos}$ violations after around $40\%$ of the total number of iterations. An intuitive explanation of this is that position violations only occur during the initial exploration phase.

\begin{table*}[]
\centering
\begin{tabular}{@{}cccc@{}}
\toprule
Task & Number of Agents & Decentralized & Centralized \\
\midrule
Swapping& $16$ & $5$ & $136$ \\
Bottleneck & $8$ & $10$ & $76$ \\
Moving obstacle & $8$ & $9$ & $36$ \\ 
Large-scale formation & $32$ & $16$ & $560$ \\
\bottomrule
\end{tabular}
\caption{Comparison of number of constraints between the centralized problem and each local subproblem of the proposed decentralized approach.}
\label{tab:constraints}
\end{table*}
\begin{table*}[]
\centering
\begin{tabular}{@{}ccccccc@{}}
\toprule
& \multicolumn{3}{c}{Max Memory Allocated (MiB)}   & \multicolumn{3}{c}{Time per Training Iteration (s)} \\
\cmidrule(lr){2-4} \cmidrule(lr){5-7}
Batch Size & Decentralized & Centralized & Percent Reduction &
Decentralized & Centralized  & Percent Reduction  \\
\midrule
32         & 1586      & 17879       & -91.13\%               & 14.75      & 1306         & -98.87\%                \\
64         & 3124      & N/A         & N/A                    & 17.92      & N/A          & N/A                     \\
384        & 18510     & N/A         & N/A                    & 38.30      & N/A          & N/A                     \\ \bottomrule
\end{tabular}
\caption{Comparison of memory usage (maximum of 10 iterations) and time per iteration (average over 10 iterations) between the decentralized and centralized formulations.
The N/A values for batch sizes $64$ and $384$ indicate that the computer ran out of memory and was unable to run a single iteration.}
\label{tab:mem}
\end{table*}
{
\newcommand{\mygriditem}[1]{%
\begin{minipage}[b]{0.24\linewidth}%
    \includegraphics[width=\textwidth]{#1}%
\end{minipage}%
}
\begin{figure*}
    \centering
    \parbox[b]{.24\linewidth}{\centering $t=0$}
    \parbox[b]{.24\linewidth}{\centering $t=1/3 T$}
    \parbox[b]{.24\linewidth}{\centering $t=2/3 T$}
    \parbox[b]{.24\linewidth}{\centering $t=T$}
    
    \vspace{0.5em}
    
        \begin{subfigure}[b]{\linewidth}
        \centering
        \mygriditem{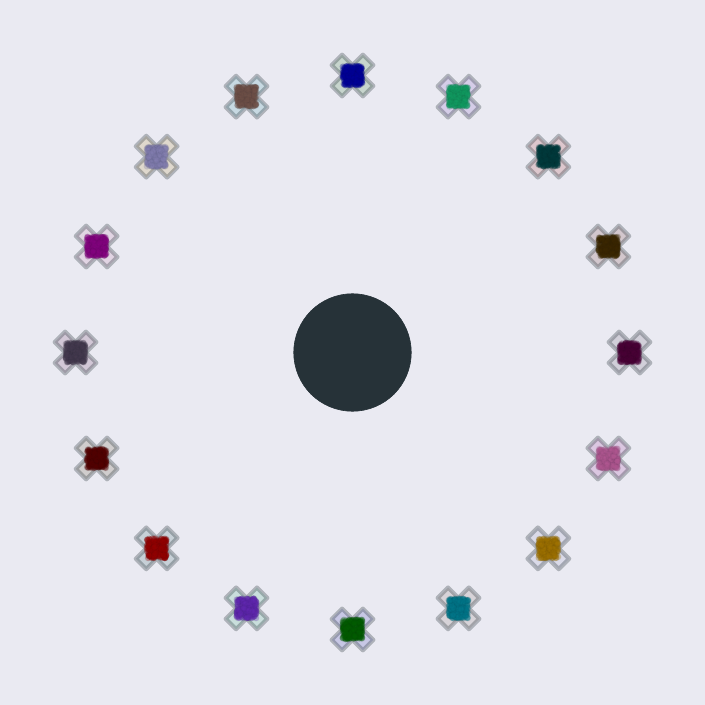}
        \mygriditem{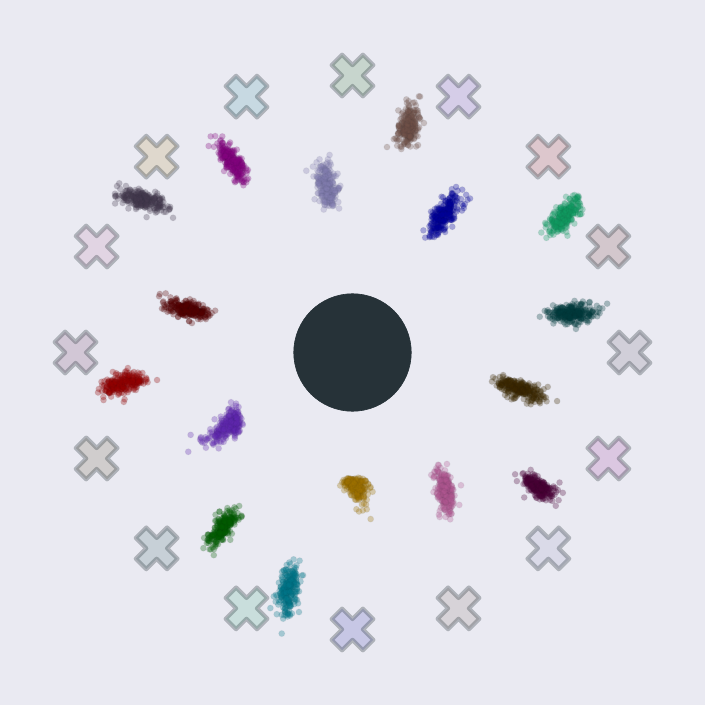}
        \mygriditem{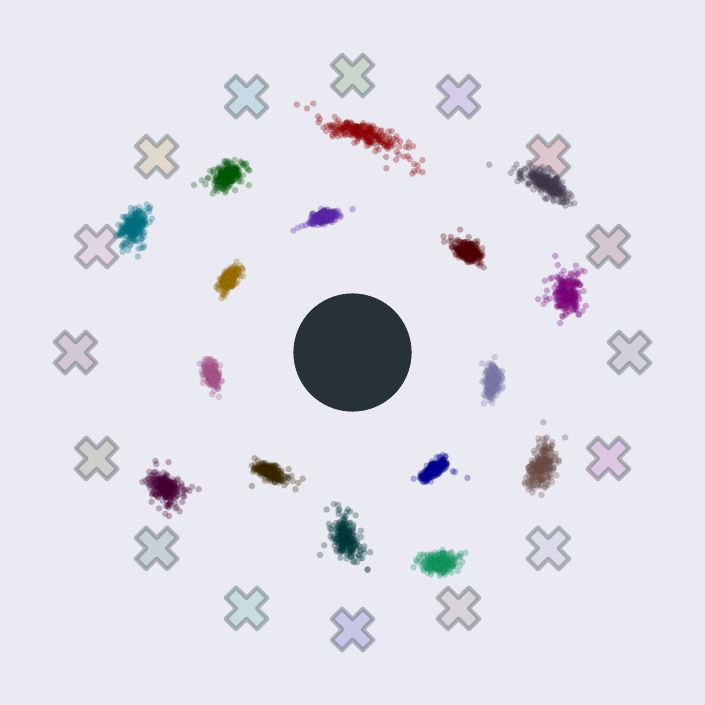}
        \mygriditem{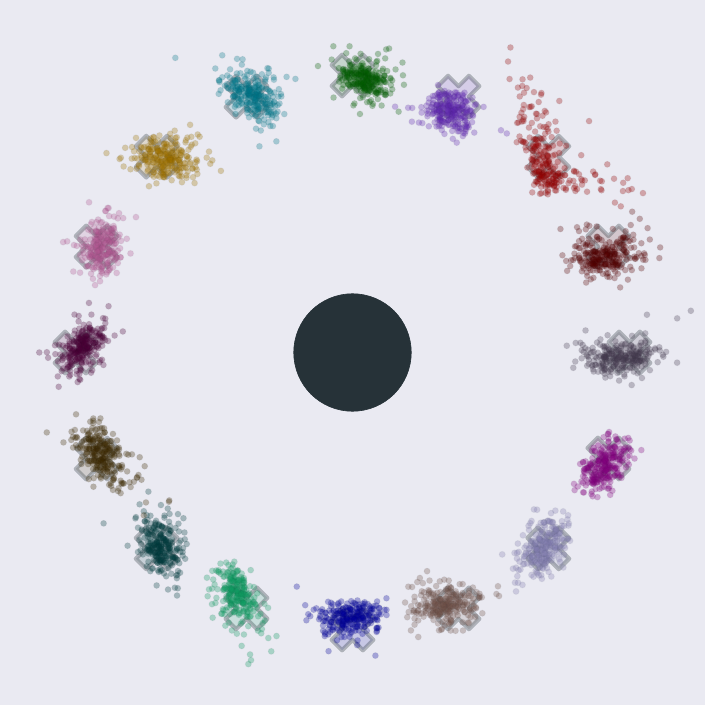}
        \caption{Symmetrical Swapping Task with One Obstacle and $16$ Agents.}
        \label{fig:exp:one_obs_swap_16}
    \end{subfigure}
    
    \vspace{0.5em}
    
    \begin{subfigure}[b]{\linewidth}
        \centering
        \mygriditem{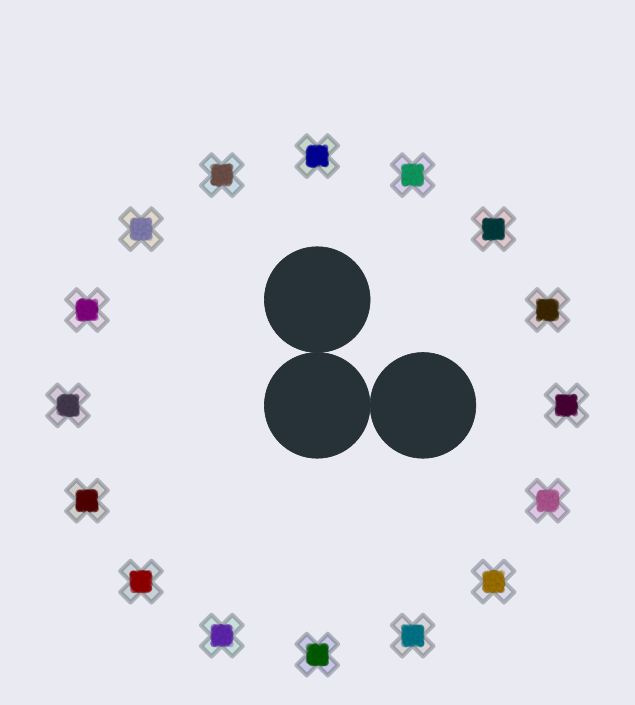}
        \mygriditem{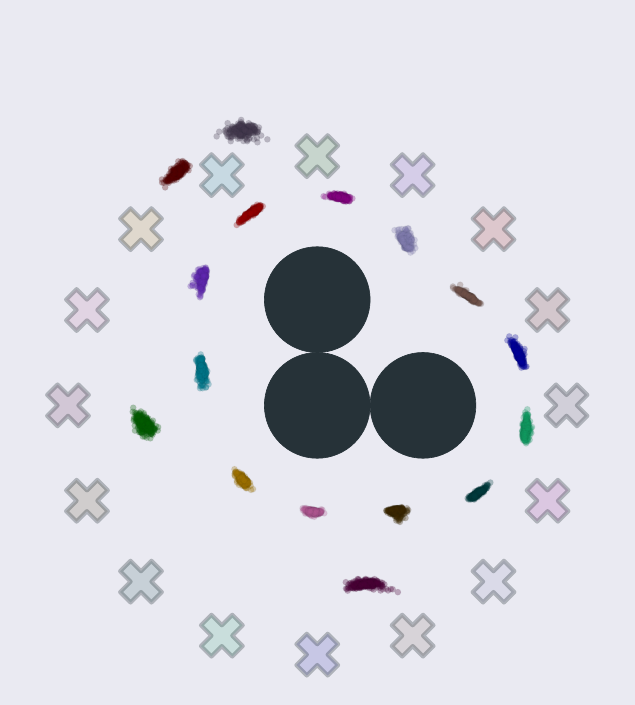}
        \mygriditem{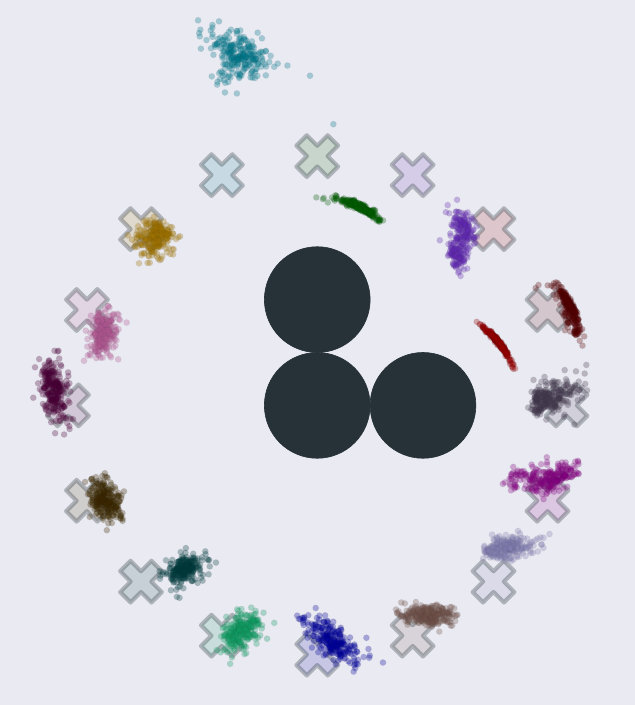}
        \mygriditem{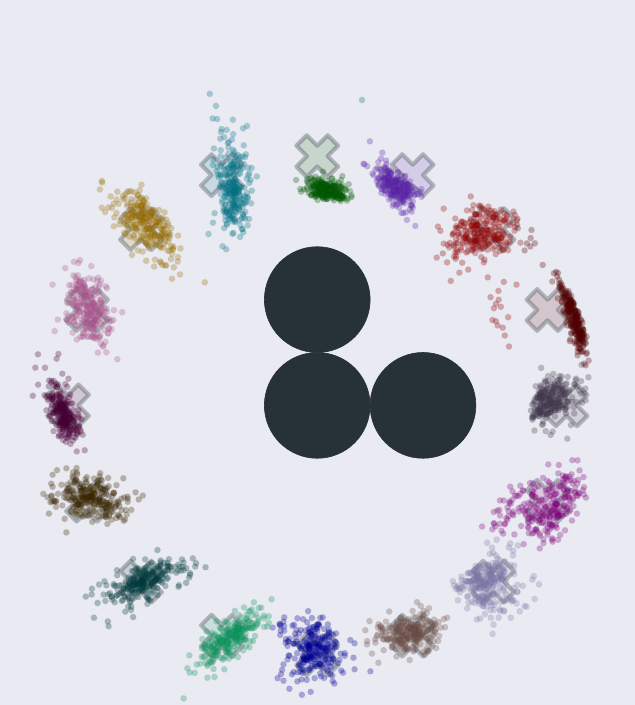}
        \caption{Unsymmetrical Swapping Task with Three Obstacles and $16$ Agents.}
        \label{fig:exp:three_obs_swap_16}
    \end{subfigure}%
    \caption{In all figures, the black circles indicate obstacles and the \textbf{X} markers indicate target positions for each agent with the same color as the marker. The snapshots demonstrate the positions of each agent for each batch at time instants $t = 0, 1/3 T, 2/3 T, T$ with the fully trained policy. The proposed approach can not only handle symmertical problems similar to \cite{pereira2020safe}, but can also successfully solve unsymmertical problems. }
    \label{fig:swap_figs}
\end{figure*}
}

{
\newcommand{\mygriditem}[1]{%
\begin{minipage}[b]{0.24\linewidth}%
    \includegraphics[width=\textwidth]{#1}%
\end{minipage}%
}
\begin{figure*}
    \centering
    \parbox[b]{.24\linewidth}{\centering $t=0$}
    \parbox[b]{.24\linewidth}{\centering $t=1/3 T$}
    \parbox[b]{.24\linewidth}{\centering $t=2/3 T$}
    \parbox[b]{.24\linewidth}{\centering $t=T$}
    
    \vspace{0.5em}
    
    % \begin{subfigure}[b]{\linewidth}
    %     \centering
    %     \mygriditem{one_obs_swap16/final_scatter_000.pdf}
    %     \mygriditem{one_obs_swap16/final_scatter_033.pdf}
    %     \mygriditem{one_obs_swap16/final_scatter_066.pdf}
    %     \mygriditem{one_obs_swap16/final_scatter_100.pdf}
    %     \caption{Swapping task with one obstacle and $16$ agents.}
    %     \label{fig:exp:one_obs_swap_16}
    % \end{subfigure}
    % 
    % \vspace{0.5em}
    
    \begin{subfigure}[b]{\linewidth}
        \centering
        \mygriditem{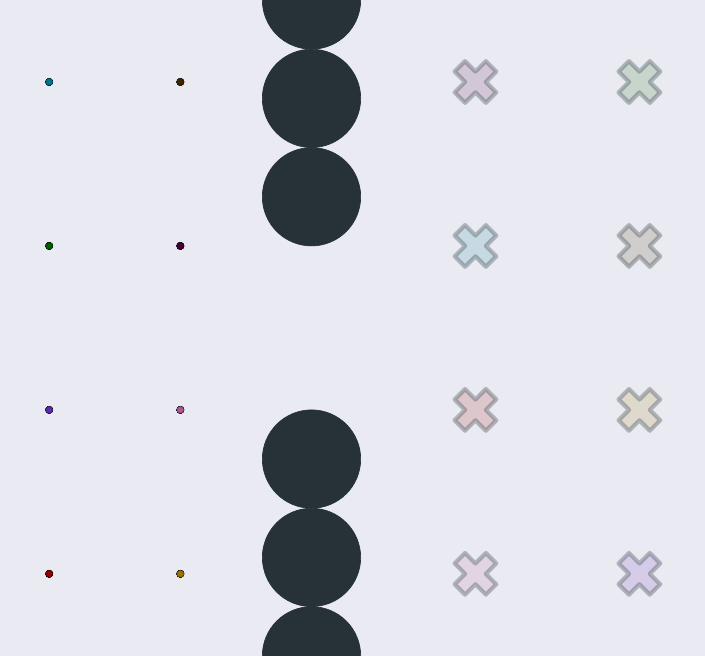}
        \mygriditem{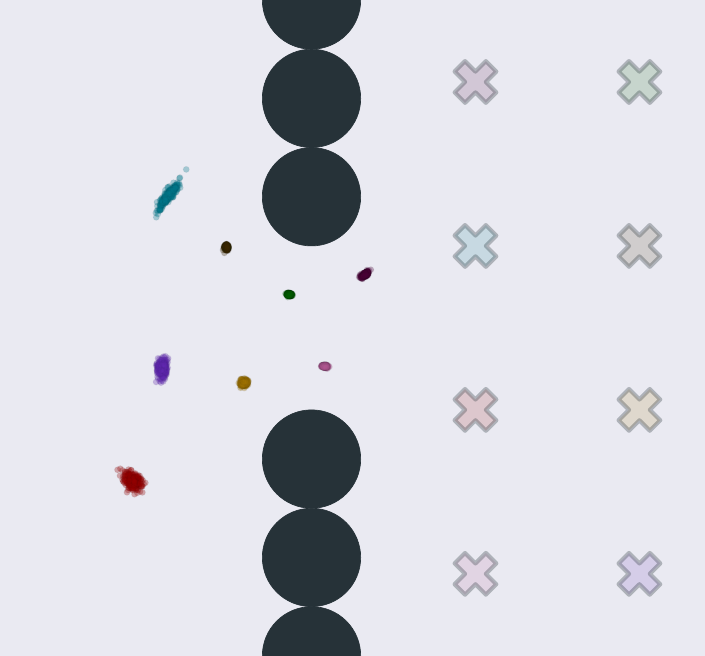}
        \mygriditem{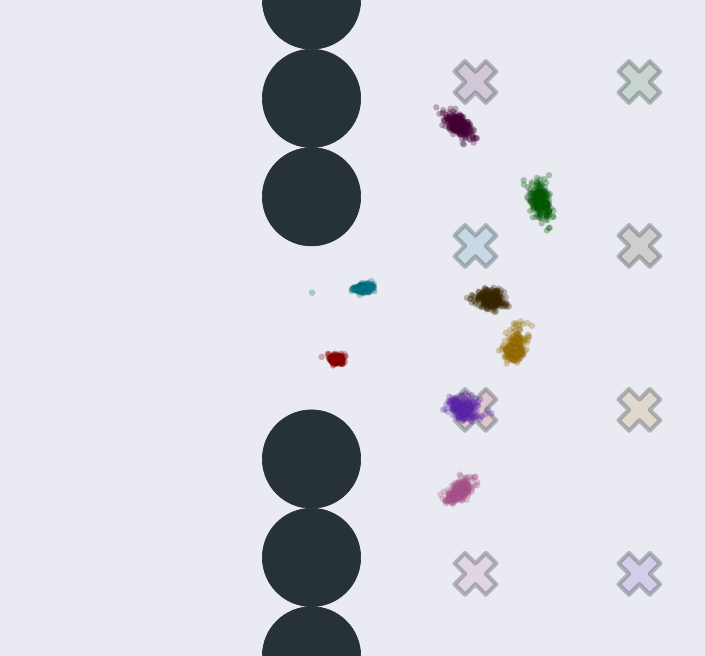}
        \mygriditem{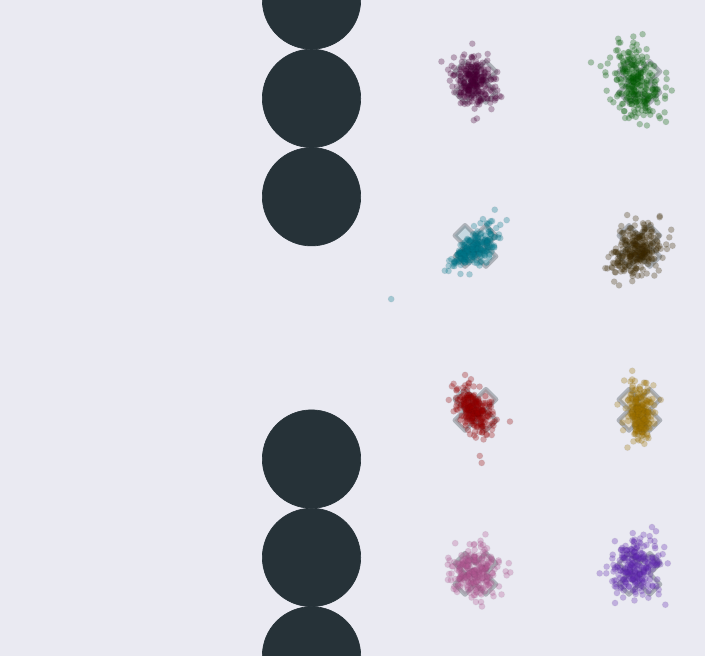}
        \caption{Bottleneck task with $8$ agents.}
        \label{fig:exp:bottleneck_8}
    \end{subfigure}
    
    \vspace{0.5em}
    
    \begin{subfigure}[b]{\linewidth}
        \centering
        \mygriditem{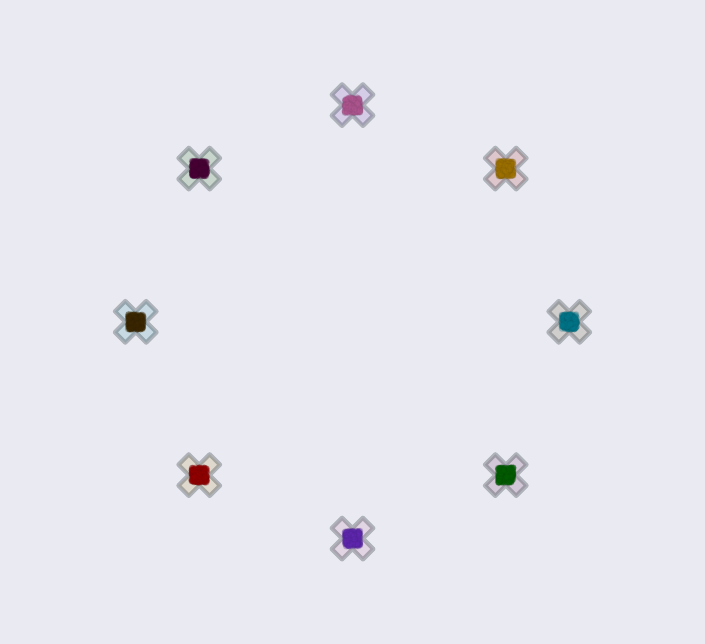}
        \mygriditem{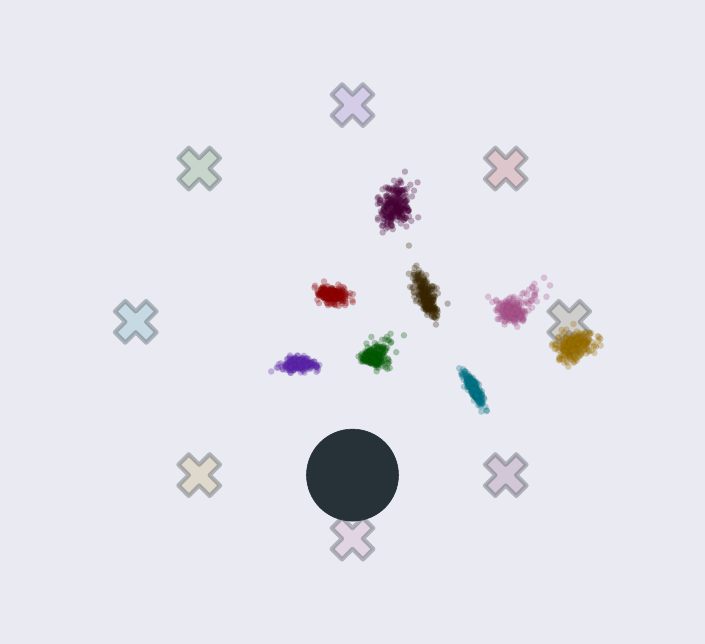}
        \mygriditem{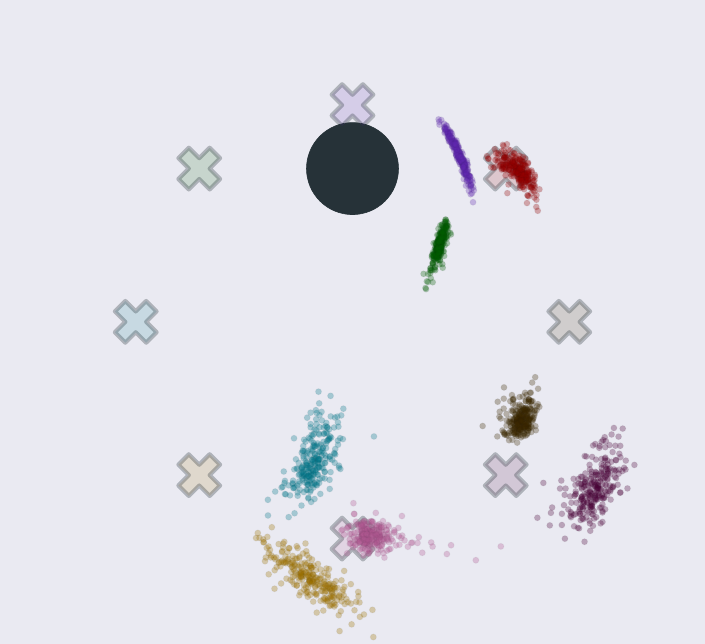}
        \mygriditem{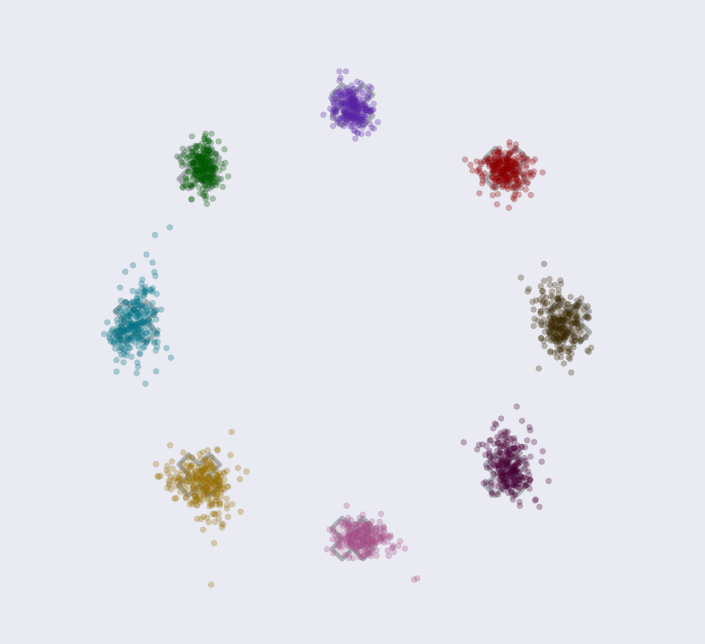}
        \caption{Swapping task with $8$ agents and a moving obstacle (uncooperative agent).}
        \label{fig:exp:moving_obs_8}
    \end{subfigure}
    
    \vspace{0.5em}
    
    \begin{subfigure}[b]{\linewidth}
        \centering
        \mygriditem{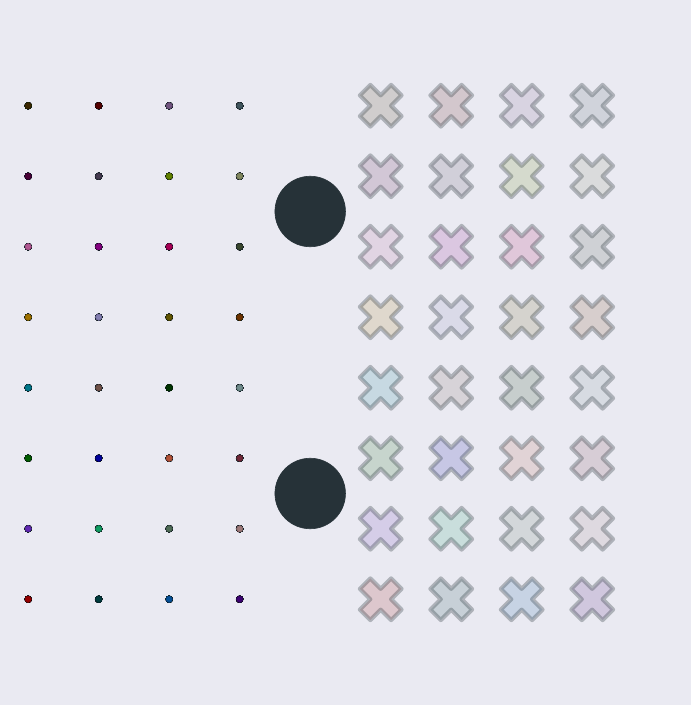}
        \mygriditem{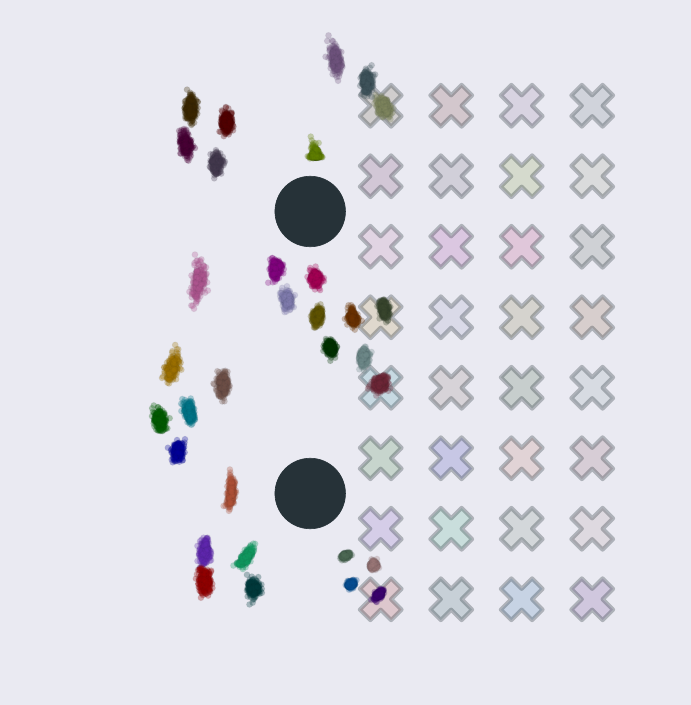}
        \mygriditem{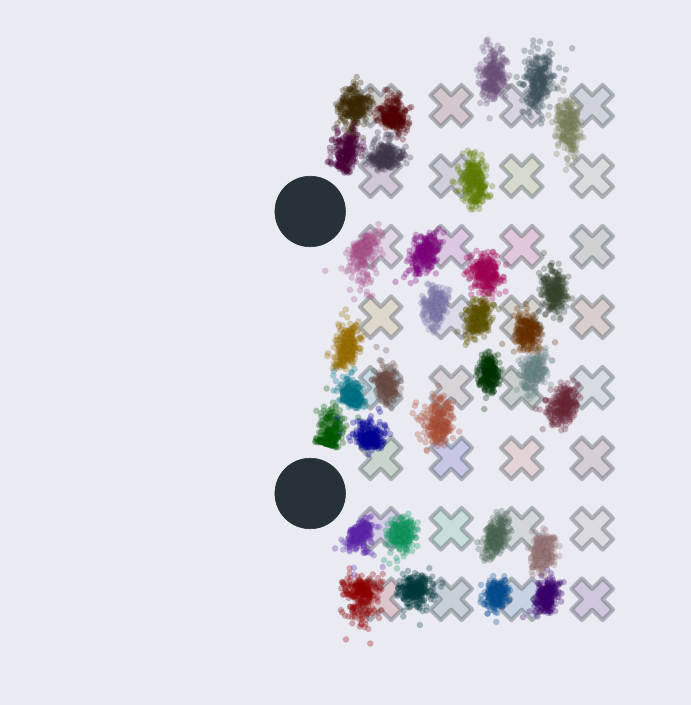}
        \mygriditem{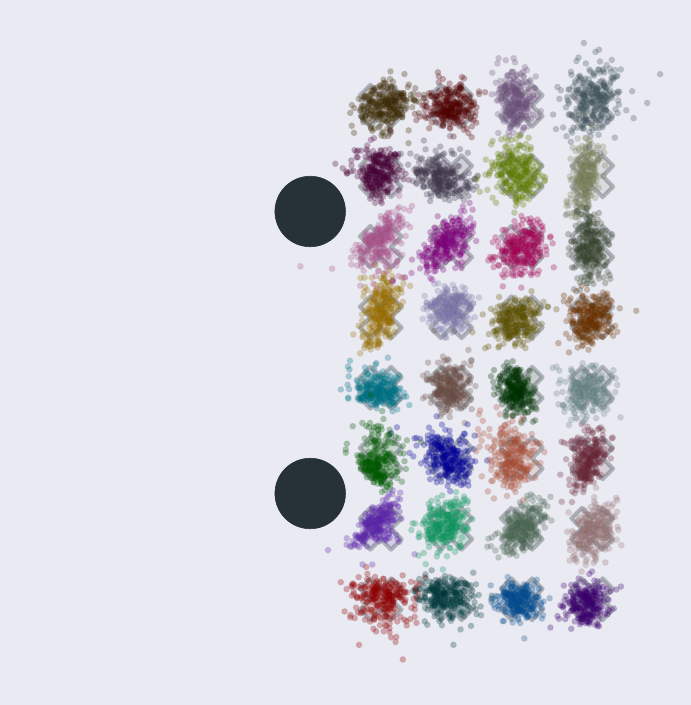}
        \caption{Large-scale formation task with $32$ agents.}
        \label{fig:exp:formation_64}
    \end{subfigure}%
    \caption{Similar to Fig. 4, black circles indicate obstacles and the \textbf{X} markers indicate target positions. The top and bottom rows involve static obstacles, while the center row considers a moving obstacle.}
    \label{fig:nonswap_figs}
\end{figure*}
}

\section{Conclusion} 
\label{sec:conclusion}
In this work, we introduced a novel and scalable deep-learning-based framework that extends the existing deep FBSDE framework to decentralized multi-agent safe stochastic optimal control problems. To achieve this, we proposed a new stochastic CBF formulation which encourages aggressive motion of moving agents, while still guaranteeing their safe operation with a high probability. Furthermore, we reformulated the multi-agent per-time-step Hamiltonian minimization problem into a decentralized version, which we solve by proposing a novel distributed ADMM-based method for large-scale quadratic optimization problems. 
% Besides safe optimal control, our proposed novel implicit layer, the Merged CADMM-OSQP layer, empowers existing implicit layers to solve QPs in a decentralized fashion, thus improving the scalability of implicit layers. 
The framework was successfully tested in simulation on several challenging multi-vehicle tasks with higher complexity and more agents as compared to previous work using a centralized approach and simpler tasks. The results demonstrate that the agents maintain their safe operation during training, thus the framework can also be appreciated from a safe reinforcement learning perspective. Finally, the scalability of the approach in terms of memory usage and training time is also validated.

\section*{Acknowledgments}
This research was supported by the ARO Award \#W911NF2010151. Augustinos Saravanos acknowledges financial support by the A. Onassis Foundation Scholarship. 

%% Use plainnat to work nicely with natbib. 

\bibliographystyle{plainnat}
\bibliography{references}

% \newpage
\section*{Supplementary Material}
In this section, we provide additional derivations and details not covered in the main paper. 
% \augustinos{Equation numbers in the supplementary material refer to
% exclusively to equations mentioned in the supplementary material. (Not needed here)}

\section{Construction of Contraint Matrix for Reduced Duplicate Centralized Problem}
\label{sec:construction_C_matrix}
For convenience of derivation and notation, it is assumed that the agents are homogeneous, i.e., all agents obey to the same dynamics with $\vx_i \in \Rb^n$ and $\vu_i \in \Rb^m$. For the same reasons, we will only assume the existence of inter-agent constraints - neglecting obstacle avoidance constraints. Nevertheless, the following derivation can be easily extended beyond these assumptions.

Let us start by restating that in the RDCP we only take into account the $N N_{\text{ineq},i} = N r$ inter-agent constraints that appear between neighboring agents out of the total $N_{\text{ineq}}$ constraints. This problem can be formulated as follows:
\begin{align}
\min_{\vu} \calH(\vu) = \sum_{i=1}^N \calH_i(\tilde{\vu}_i) \quad \text{s.t. } \vC \vu \leq \vd 
\label{Reduced Dup Centralized Problem}
\end{align}
where $\calH(\vu) = \frac{1}{2} \vu \T \vR \vu + \vq\T \vu$
with $\vq \coloneqq \vG\T \frac{\partial V}{\partial \vx}$. Regarding the construction of matrix $\vC$, let us first express all $\vA_i \in \Rb^{r \times (r+1) m}$, $i \in \llbracket 1, N \rrbracket$ as block matrices consisting of $r \times (r+1)$ blocks as follows:
\begin{equation}
\vA_i = 
\begin{bmatrix}
\va_i^{1, 1} & \va_i^{1, 2} & \mathbf{0} & \dots & \mathbf{0} \\
\va_i^{2, 1} & \mathbf{0} & \va_i^{2, 2} & \dots & \mathbf{0} \\
\vdots & \vdots & \vdots & \ddots & \vdots \\ 
\va_i^{r, 1} & \mathbf{0} & \mathbf{0} & \dots & \va_i^{r, 2}
\end{bmatrix}
\end{equation}
where each block $\va_i^{k,b} \in \Rb^{m}$, $k = \llbracket 1, r \rrbracket, \ b = 1,2$  is a row vector. The blocks $\va_i^{k,1}$ are the ones always multiplied with the ego part $\vu_i$ of $\tilde{\vu}_i$, while the blocks $\va_i^{k,2}$ correspond to the neighbors parts $\vu_j^{(i)}, \ j \in \calN_i$ of $\tilde{\vu}_i$.
Based on that, $\vC = [\vC_1 ; \dots ; \vC_N]$, where each $\vC_i \in \Rb^{r \times N m}$ can be viewed as a block matrix consisting of $r \times N$ blocks of size $\Rb^m$, and is constructed as follows: for all $k = \llbracket 1, r \rrbracket$, i) each $\vA_i[k, 1]$ (i.e., $\va_i^{k,1}$) is assigned to block $\vC_i[k, i]$ and ii) each $\vA_i[k, l]$ (i.e., $\va_i^{k,2}$) to $\vC_i[k, j]$ with $l \in \llbracket 2, r+1 \rrbracket$. 

\section{Proof of Lemmas} \label{sec:lemmas_proofs_section}
Let $\mathcal{D} \coloneqq \{ \mathcal{D}_l \}_{l=1}^\Nineqb$
with $\mathcal{D}_l \coloneqq \{ \vcdij \}_{j=1}^{\xi_l}$
denote the set of equivalence classes induced by the equality equivalence relation
\begin{equation}
    \vc_{\zeta_{l,a}} \sim \vc_{\zeta_{l,b}} \iff
    \vc_{\zeta_{l,a}} = \vc_{\zeta_{l,b}}, \quad \forall a, b \in \llbracket 1, \xi_l \rrbracket \;
        \forall l
\end{equation}
where $\zeta_{l, j}$ denotes the row corresponding to the $j$-th instance of the $l$-th unique constraint.
In other words, there are $\xi_l$ duplicates for each of the $\Nineqb$ unique constraint rows.

% \begin{assumption}
% \label{assumption:lemma_assumption}
% Suppose that the control cost matrix, $\vR$, is positive definite and LICQ holds for the reduced non-duplicate problem.
% \end{assumption}

% Before the proof of Lemma 1, we start with providing a useful lemma that allows us to only consider the active constraints in a KKT system.

We will now proceed with proving Lemma 1, which we restate for the convenience of the reader.
\begin{customlemma}{1}
\label{lemma:lemma_1}
Given that Assumption \ref{assumption:lemma_assumption} holds, then it follows that,
\begin{align}
    \lambdanondup{l} &= \sum_{j=1}^{\xi_l} \lambdadup{\zeta_{l,j}} \quad \forall l \in \llbracket 1, \Nineqb  \rrbracket
    \label{eq:relation_dup:sum_lambda} 
    % \\
    % \lamdlamnondup{i} &= \sum_{j=1}^{\xi_i} \lamdlamdup{\zeta_{i,j}}
    % \label{eq:relation_dup:sum_lambdadlambda}
\end{align}
where $\blambdadup, \blambdanondup$ denote the vectors of Lagrange multipliers 
% and $\dblambdadup, \dblambdanondup$ denote the variables of the KKT system 
for the duplicate and non-duplicate problems respectively.
% \\\marcus{in the main paper we use different symbols for parameters e.g., d instead of zeta. We should adopt these in the main paper}
\end{customlemma}
% --------------------------------------------------
\begin{proof}
Let $\zeta_l \coloneqq \zeta_{l,1}$ denote a representative
of the $l$-th equivalence class, where
$c_{\zeta_l} = c_{\zeta_{l, j}}$ for all $l, j$.
From the definition of $\zeta_{l,j}$,
we can represent the non-duplicate
$\bar{\vC}$ and $\bar{\vd}$ as,
\begin{align}
    \bar{\vC} &= [ \vc_{\zeta_1}; \dots; \vc_{\zeta_\Nineqb} ]
    \implies \bar{\vc}_l = \vc_{\zeta_l} \label{eq:rel_c_cbar} \\
    \bar{\vd} &= [ \vd_{\zeta_1}; \dots; \vd_{\zeta_\Nineqb} ]
    \implies \bar{\vd}_l = \vd_{\zeta_l} \label{eq:rel_d_dbar}.
\end{align}
As we do not assume that agents will be mutual neighbors of each other, note that $\mathcal{D}$ may contain singleton sets, i.e., there may be constraints rows that are unique in $\vC$.

The Lagrangian for the duplicate problem is given by,
\begin{equation}
    \mathcal{L}_{\text{dup}} = \calH(\vu) + \blambdadup\T (\vC \vu - \vd)
\end{equation}
where we remind the reader that matrices for the duplicate problem have dimension $\vC \in \Rb^{N r \times N m}$ and $\vd \in \Rb^{N r}$, where based on \Cref{ass:fixed_neighborhood}.
Since the constraint rows of each equivalence class are identical,
by representing the matrix multiplication with $\vC\T$ as a sum over the rows of $\vC$ we can factor out $\vcdi$ over each equivalence class, yielding
\begin{align}
    \phantom{{}={}} \mathcal{L}_{\text{dup}} &= \calH(\vu) + \blambdadup\T (\vC \vu - \vd) \\
    &= \calH(\vu) + \sum_{k=1}^{N r} \lambdadup{k} \left(\vc_k \vu   - d_k\right)
    \\
    &= \calH(\vu) + \left(\sum_{k=1}^{N r} \lambdadup{k} \vc_k\right) \vu
- \sum_{k=1}^{N r} \lambdadup{k} \, d_k 
\end{align}
\begin{align}
    \begin{split}
    &= \calH(\vu) + \left(\sum_{l=1}^\Nineqb \left(\sum_{j=1}^{\xi_l} \lambdadup{\zeta_{l,j}} \right) \vc_{\zeta_l} \right) \vu \\
    &\quad - \sum_{l=1}^\Nineqb \left( \sum_{j=1}^{\xi_l} \lambdadup{\zeta_{l,j}} \right) d_{\zeta_l}
    \end{split} \\
    &= \calH(\vu) + \left(\sum_{l=1}^\Nineqb \nu_l \bar{\vc}_l \right) \vu
    - \sum_{l=1}^\Nineqb \nu_l \bar{d}_l \\
    &= \calH(\vu) + \bnu\T (\bar{\vC} \vu - \bar{\vd}) \label{eq:dup_lagrange}
\end{align}
where we have defined the vector $\bnu$ such that
\begin{equation}
    \nu_l = \sum_{j=1}^{\xi_l} \lambdadup{\zeta_{l,j}}.
\end{equation}
On the other hand, the Lagrangian for the non-duplicate problem reads,
\begin{equation} \label{eq:nondup_lagrange}
    \mathcal{L}_{\text{non-dup}} = \calH(\vu) + \blambdanondup\T (\bar{\vC} \vu - \bar{\vd}).
\end{equation}
We now assume that the control cost matrix, $\vR$, is positive definite for our problem which is a standard assumption for stochastic optimal control.
Under LICQ, the Lagrange multipliers corresponding to the saddle point of $\mathcal{L}_{\text{non-dup}}$ are unique.
Thus, comparing \eqref{eq:dup_lagrange} and \eqref{eq:nondup_lagrange}, we must have that $\bnu = \blambdanondup$, i.e.,
\begin{equation}
    \lambdanondup{l} = \nu_l = \sum_{j=1}^{\xi_l} \lambdadup{\zeta_{l,j}} \quad \forall l \in \llbracket 1, \Nineqb  \rrbracket.
\end{equation}
Thus, \eqref{eq:relation_dup:sum_lambda} holds.
\end{proof}

\begin{customlemma}{2}[QP Gradients]
\label{lemma:lemma_2}
Let $\vu^*$ and $\blambdadup^*$ denote the solutions to the reduced duplicate problem at any given time step $t$, 
\begin{align}
    \min_{\vu(t)} &\quad \frac{1}{2} \vu(t)\T \vR(t) \vu(t) + \vq(t)\T \vu(t) \\
    \textrm{s.t.} &\quad \vC(t) \vu(t) \leq \vd(t)
\end{align}
and let $\ell$ denote the overall neural-network training loss function.
Then, the gradients of $\ell$ with respect to the data matrices $\vR, \vq, \vC, \vd$ for the time step $t$ have the form,
\begin{subequations}
\begin{align}
\nabla_\vq \ell  &= \dvu
\label{eq:grad_q} \\
\nabla_{\vd} \ell  &= -\diag(\blambdadup^*) \dblambdadup
\label{eq:grad_d} \\
\nabla_\vR \ell &= \frac{1}{2}(\dvu \, {\vu^*}\T + \vu^* \, \dvu\T)
\label{eq:grad_R} \\
\nabla_{\vC} \ell &= \blambdadup^* \, \dvu\T + \diag(\blambdadup^*) \dblambdanondup {\vu^*}\T 
\label{eq:grad_C}
\end{align}
\end{subequations}
where $\dvu$ and $\dblambdanondup$ are the solutions to the KKT system
\begin{equation}
    \begin{bmatrix}
    \vR & \vC\T \diag(\blambdadup^*) \\
    \vC & \diag(\vC \vu^* - \vd)
    \end{bmatrix}
    \begin{bmatrix}
    \dvu \\ \dblambdadup
    \end{bmatrix}
    =
    \begin{bmatrix} -\nabla_{\vu^*} \ell \\ \mathbf{0} \end{bmatrix}.
\end{equation}
\end{customlemma}

\begin{proof}
The KKT conditions for the reduced duplicate problem are given by,
\begin{subequations}
\begin{align}
\vR \vu^* + \vq + \vC \T \blambda_{\text{dup}}^* & = \mathbf{0}
\\ 
\diag(\blambda_{\text{dup}}^*) (\vC \vu^* - \vd) & = \mathbf{0}
\\ 
\blambda_{\text{dup}}^* & \geq \mathbf{0}
\\ 
\vC \vu^* & \leq \vd .
\end{align}
\end{subequations}
Next, let us define:
\begin{equation}
\bepsilon(\bdelta, \theta) = 
\begin{bmatrix}
\vR \vu^* + \vq + \vC \T \blambda_{\text{dup}}^* 
\\ 
\diag(\blambda_{\text{dup}}^*) (\vC \vu^* - \vd) 
\end{bmatrix}
\label{KKT epsilon}
\end{equation}
where $\bdelta = [\vu^*;\,\blambda_{\text{dup}}^*]$ and $\theta$ denotes any parameter in \eqref{KKT epsilon}, i.e., $\vR, \vq, \vC, \vd$.
To obtain the gradient of the training loss $\ell$ with respect to the parameters $\theta$, we have that,
\begin{equation}
    \nabla_\theta \ell
    = (\nabla_\theta \bdelta) \T (\nabla_{\bdelta} \vu^*) \T \nabla_{\vu^*} \ell
\end{equation}
which leads to,
\begin{align}
(\nabla_\theta \ell) \T
&= (\nabla_{\vu^*} \ell)\T \nabla_{\bdelta} \vu^* \nabla_\theta \bdelta \nonumber\\
&= (\nabla_{\vu^*} \ell)\T \nabla_{\bdelta} \vu^* \underbrace{\big(-\nabla_{\bdelta} \bepsilon(\bdelta, \theta)^{-1} \nabla_\theta \bepsilon(\bdelta, \theta) \big)}_{\text{using the Implicit Function Theorem}}
\label{chain rule}
\end{align} 
Next, we write the above equation as follows,
\begin{equation} \label{eq:implicit_gradient}
\nabla_\theta \ell \T = \bomega \, \nabla_\theta \bepsilon(\bdelta, \theta)
\end{equation}
where we define $\bomega$ as,
\begin{equation}
\bomega 
= 
\begin{bmatrix} 
\dtot \vu \T & \dtot \blambda_{\text{dup}} \T 
\end{bmatrix} 
= 
-(\nabla_{\vu^*} \ell)\T (\nabla_{\bdelta} \vu^*) \nabla_{\bdelta} \bepsilon(\bdelta, \theta)^{-1}.
\end{equation}
This is equivalent to solving the following linear system of equations for $\bomega$,
\begin{align}
\bomega \nabla_{\bdelta} \bepsilon(\bdelta,\, \theta)
&= - (\nabla_{\vu^*} \ell)\T \,\,\nabla_{\bdelta} \vu^* 
\nonumber
\\
&= - (\nabla_{\vu^*} \ell)\T \,\, 
\begin{bmatrix}
\vI & \mathbf{0} 
\end{bmatrix}
\nonumber
\\
&= -
\begin{bmatrix}
(\nabla_{\vu^*} \ell)\T & \mathbf{0} \end{bmatrix} .
\label{omega eq}
\end{align}
Now, given that,
\begin{equation}
\nabla_{\bdelta} \bepsilon(\bdelta,\, \theta) = 
\begin{bmatrix}
\vR & \vC \T \\ 
\diag(\blambda_{\text{dup}}^*) \vC & \diag(\vC \vu^* - \vd)
\end{bmatrix}
\end{equation}
taking the transpose of both sides in \eqref{omega eq}, we obtain,
\begin{equation}
\begin{bmatrix}
\vR & \vC \T \diag(\blambdadup^*)  \\[0.2em]
\vC & \diag(\vC \vu^* - \vd)
\end{bmatrix}
\begin{bmatrix}
\dvu \\ \dblambdadup
\end{bmatrix}
=
\begin{bmatrix}
-\nabla_{\vu^*} \ell \\ 
\mathbf{0}
\end{bmatrix}.
\label{eqn:dup_kkt_system}
\end{equation}
Having solved for $\bomega$ with \eqref{eqn:dup_kkt_system},
we first compute the gradients of $\ell$ w.r.t the vectors $\vq$ and $\vd$ by using \eqref{eq:implicit_gradient},
\begin{align}
    \nabla_\vq \ell\T &= \bomega \, \nabla_\vq \bepsilon
        = \begin{bmatrix}\dvu\T & \dblambdadup\T \end{bmatrix}
        \begin{bmatrix}\vI \\ 0 \end{bmatrix}= \dvu\T \nonumber \\
        \therefore \nabla_\vq \ell &= \dvu \\
    \nabla_\vd \ell\T &= \bomega \, \nabla_\vd \bepsilon
        = \begin{bmatrix}\dvu\T & \dblambdadup\T \end{bmatrix}
        \begin{bmatrix}0 \\ -\diag(\blambdadup^*) \end{bmatrix} \nonumber \\
    \therefore \nabla_\vd \ell  &= -\diag(\blambdadup^*) \dblambdadup .
\end{align}
For the matrix parameters $\vR, \vC$, note that, for example, $\nabla_\vR \bepsilon$ is a three-dimensional tensor.
To avoid tensor operations, we will instead use the following relation to solve for the gradient,
\begin{equation} 
    \df(X) = \tr\Big[ \nabla_\vX f(\vX)\T \dvX \Big]
\end{equation}
for any function $f : \Rb^{n \times n} \to \Rb$, where $\df$ and $\dvX$ denote the differential of $f$ and $\vX$ respectively.
For $\vR$, we get,
\begin{align}
\dbepsilon
    &= \begin{bmatrix} \dvR \vu^* \\ \mathbf{0} \end{bmatrix} \nonumber \\
\implies
\dell
    &= \begin{bmatrix} \dvu\T & \dblambdadup\T \end{bmatrix} \begin{bmatrix} \dvR \vu^* \\ 0 \end{bmatrix} \nonumber \\
    &= \dvu\T \dvR \, \vu^* \nonumber \\
    &= \frac{1}{2} \Big( \dvu\T \dvR \, \vu^* + {\vu^*}\T \dvR \, \dvu \Big) \nonumber\\
    &= \frac{1}{2} \tr\Big( (\vu^* \dvu\T + \dvu \, {\vu^*}\T) \dvR \Big) \nonumber\\
    &= \tr\Big( \frac{1}{2}(\dvu \, {\vu^*}\T + \vu^* \, \dvu\T)\T \dvR \Big) \nonumber \\
\implies
\nabla_\vR \ell &= \frac{1}{2}(\dvu \, {\vu^*}\T + \vu^* \, \dvu\T) .
\end{align}
Similarly for $\vC$, we have,
\begin{align}
\dbepsilon &= \begin{bmatrix} \dvC\T \blambdadup^* \\ \diag(\blambdadup^*) \dvC \vu^* \end{bmatrix} \nonumber \\
\implies
\dell
&= \begin{bmatrix} \dvu\T & \dblambdadup\T \end{bmatrix} 
    \begin{bmatrix} \dvC\T \blambdadup^* \\ \diag(\blambdadup^*) \dvC \vu^* \end{bmatrix} \nonumber \\
&= \dvu\T \dvC\T \blambdadup^* + \dblambdadup\T \diag(\blambdadup^*) \dvC \vu^* \nonumber \\
% &= \tr \Big[ \blambdadup\T \dvC \, \dvu + \vu \, \dblambdadup\T \diag(\blambdadup) \dvC \Big] \\
    % &= \tr \Big[ \left(\dvu \, \blambdadup\T + \vu \, \dblambdadup\T \diag(\blambdadup) \right) \dvC \Big] \\
&= \tr \Big[ \left(\blambdadup^* \, \dvu\T + \diag(\blambdadup^*) \dblambdadup {\vu^*}\T \right)\T \dvC \Big] \nonumber \\
\implies
\nabla_\vC \ell &= \blambdadup^* \, \dvu\T + \diag(\blambdadup^*) \dblambdadup {\vu^*}\T .
\end{align}
\end{proof}

\begin{customlemma}{3}
\label{lemma:lemma_3}
Given that Assumption \ref{assumption:lemma_assumption} holds, then it follows that,
\begin{align}
    \lamdlamnondup{l} &= \sum_{j=1}^{\xi_l} \lamdlamdup{\zeta_{l,j}},\, \forall l \in \llbracket 1, \Nineqb  \rrbracket
    \label{eq:relation_dup:sum_lambdadlambda}
\end{align}
where 
% $\blambdadup, \blambdanondup$ denote the Lagrange multipliers 
$\dblambdadup, \dblambdanondup$ denote the vectors of the variables of the KKT systems for the duplicate and non-duplicate problems respectively.
% \\\marcus{in the main paper we use different symbols for parameters e.g., d instead of zeta. We should adopt these in the main paper}
\end{customlemma}

\begin{proof} We drop the $*$ from $\vu^*$ and $\blambdadup^*$ for ease of notation. 
To show that that \eqref{eq:relation_dup:sum_lambdadlambda} holds,
consider the first row-block of the KKT system for
the duplicate problem 
\begin{equation}
    \vR \mathrm{d}\vu + \vC\T\diag(\blambda_\text{dup})\mathrm{d}\blambda_\text{dup} = -\nabla_\vu \ell .
    \label{eqn:dup_first_block}
\end{equation}
Representing the matrix multiplication with $\vC\T$ as a sum over the rows of $\vC$, we can factor out $\vcdi$ over each equivalence class, yielding,
\begin{align}
    \vC\T\diag(\blambdadup) \dblambdadup
    &= \sum_{k=1}^{Nr} \lamdlamdup{k} \vc_k\T \\
    &= \sum_{l=1}^\Nineqb \sum_{j=1}^{\xi_l} \lamdlamdup{\zeta_{l,j}} \vcdij\T \\
    &= \sum_{l=1}^\Nineqb \left( \sum_{j=1}^{\xi_l} \lamdlamdup{\zeta_{l,j}} \right) \vcdi\T \\
    &= \sum_{l=1}^\Nineqb \chi_l \vcdi\T \\
    &= \bar{\vC}\T \bchi \label{eq:dup_first_row_block}
\end{align}
where we have defined the vector $\bchi \in \Rb^{\Nineqb}$ such that
\begin{equation}
    \chi_l \coloneqq \sum_{j=1}^{\xi_l} \lamdlamdup{\zeta_{l,j}}.
\end{equation}

The KKT system for the duplicate problem reads
\begin{equation}
    \begin{bmatrix}
        \vR & \vC\T \diag(\blambdadup) \\
        \vC & \diag(\vC \vu - \vd)
    \end{bmatrix}
    \begin{bmatrix}
        \dvu \\
        \dblambdadup
    \end{bmatrix}
    =
    \begin{bmatrix}
        -\nabla_\vu \ell \\
        \vzero
    \end{bmatrix}.
\end{equation}
In order to eliminate the
KKT variables related to the inactive constraints (i.e., $\mathrm{d}\blambda_\text{dup,inactive}$), we need to first rewrite the above KKT matrix. For this, let $\vC_\cactive$ and $\vC_\cinactive$ denote the blocks of the matrix $\vC$ representing rows of active and inactive constraints. Then w.l.o.g., we can assume that the decision variables $\mathrm{d}\vu$ are sorted in such a way that the KKT matrix above can be rewritten as follows,
\begin{equation}
    \begin{bmatrix}
        \vR & \vC_\cactive \T \diag(\blambda_\cactive) & \vC_\cinactive\T \diag(\blambda_\cinactive) \\
        \vC_\cactive & \diag(\vC_\cactive \vu - \vd_\cactive) & 0 \\
        \vC_\cinactive & 0 & \diag(\vC_\cinactive \vu - \vd_\cinactive)
    \end{bmatrix}
\end{equation}
The Schur complement of the bottom right block 
% (i.e., the KKT matrix from performing block elimination on the inactive variables) 
given by,
\begin{equation}
    \begin{bmatrix}
        \vR & \vC_{\cactive}\T \diag(\blambda_\cactive) \\
        \vC_\cactive & 0
    \end{bmatrix}
\end{equation}
is easily derived by applying complementary slackness to $\blambda_\cinactive$
    to observe that $\vC_\cinactive\T \diag(\blambda_\cinactive) = 0$ and $\diag(\vC_\cactive \vu - \vd_\cactive) = 0$.\\
Next, we can eliminate $\mathrm{d}\blambda_\text{dup,inactive}$ and solve for the remaining KKT variables using the Schur complement as follows,
\begin{equation}
    \begin{bmatrix}
        \vR & \vC_\cactive\T \diag(\blambdadupactive) \\
        \vC_\cactive &  \vzero
    \end{bmatrix}
    \begin{bmatrix}
        \dvu \\
        \dblambdadupactive
    \end{bmatrix}
    =
    \begin{bmatrix}
        -\nabla_\vu \ell \\
        \vzero
    \end{bmatrix}.
\end{equation}
Since the (2,2) block of the Schur complement is zero, noting \eqref{eq:dup_first_row_block}, we perform a change of variable with $\bchi$ to give
\begin{equation} \label{eq:dup_reduced}
    \begin{bmatrix}
        \vR & \bar{\vC}_\cactive\T \\
        \bar{\vC}_\cactive & 0
    \end{bmatrix}
    \begin{bmatrix}
        \dvu \\
        \chi_\cactive
    \end{bmatrix}
    =
    \begin{bmatrix}
        -\nabla_\vu \ell \\
        \vzero
    \end{bmatrix}.
\end{equation}

We now connect this with the KKT system of the non-duplicate problem, which reads
\begin{equation}
    \begin{bmatrix}
        \vR & \bar{\vC}\T \diag(\blambda) \\
        \bar{\vC} & \diag(\bar{\vC} \vu - \vd)
    \end{bmatrix}
    \begin{bmatrix}
        \dvu \\
        \dblambdanondup
    \end{bmatrix}
    =
    \begin{bmatrix}
        -\nabla_\vu \ell \\
        \vzero
    \end{bmatrix}.
\end{equation}
Similar to the duplicate problem, we can use the Schur complement to eliminate the KKT variables corresponding to the inactive constraints for the non-duplicate problem giving,
\begin{equation}
    \begin{bmatrix}
        \vR & \vC_\cactive\T \diag(\blambdanondupactive) \\
        \vC_\cactive & 0
    \end{bmatrix}
    \begin{bmatrix}
        \dvu \\
        \dblambdanondupactive
    \end{bmatrix}
    =
    \begin{bmatrix}
        -\nabla_\vu \ell \\
        \vzero
    \end{bmatrix}.
\end{equation}
Define $\bgamma \coloneqq \diag(\blambdanondup) \dblambdanondup$.
Performing a change of variables to consider
$\bgamma_\cactive$ instead yields
\begin{equation} \label{eq:nondup_reduced}
    \begin{bmatrix}
        \vR & \vC_\cactive\T \\
        \vC_\cactive & 0
    \end{bmatrix}
    \begin{bmatrix}
        \dvu \\
        \bgamma_\cactive
    \end{bmatrix}
    =
    \begin{bmatrix}
        -\nabla_\vu \ell \\
        \vzero
    \end{bmatrix}.
\end{equation}
Consequently,
we see that the KKT matrices in \eqref{eq:dup_reduced} and \eqref{eq:nondup_reduced} are equivalent.
Since $\vR$ is positive definite and hence invertible, invoking LICQ gives us that the KKT matrix is also invertible.
Hence, the unique solution to both \eqref{eq:dup_reduced} and \eqref{eq:nondup_reduced} are equivalent,
i.e., $\dvu$ and the active rows of $\bchi_\cactive$ and $\bgamma_\cactive$ must coincide.

For the inactive rows of $\bchi$ and $\bgamma$, note that
both $\blambdadupinactive$ and $\blambdanondupinactive$ are zero by complementary slackness.
Consequently,
\begin{equation}
    \bgamma_{\cinactive} = \bchi_{\cinactive} = 0.
\end{equation}
Hence, $\bgamma = \bchi$.
In other words,
\begin{equation}
    \lamdlamnondup{l} = \sum_{j=1}^{\xi_l} \lamdlamdup{\zeta_{l,j}},\quad \forall l \in \llbracket 1, \Nineqb  \rrbracket,
\end{equation}
and thus \eqref{eq:relation_dup:sum_lambdadlambda} holds.
\end{proof}

\section{Proof of Theorem 1}
\label{sec:theorem_proof_section}
% To prove Theorem 1, we first prove the following lemma which
% shows the form of the gradient updates for the non-duplicate problem.
We are now ready to prove Theorem 1, which we restate below
for the convenience of the reader.
%%%%%%%%%%%%%%%%%%%%%%%%%%%%%%%%%%%%%%%%%%%%%%%%%%%%%%%%%%%%%%%%%%
\begin{customthm}{1} \label{thm:same_gradient}
Let $\vM$ denote the matrix describing the relationship between the duplicate and non-duplicate constraints:
\begin{equation}
  \vC = \vM \bar{\vC}, \quad \vd = \vM \bar{\vd} \label{eq:M_def}
\end{equation}
Then, for loss $\ell$, the gradients $\nabla_\vR \ell, \nabla_\vq \ell, \nabla_\vC \ell, \nabla_\vd \ell$
coincide for the duplicate and non-duplicate problems and are unique.
\end{customthm}
%%%%%%%%%%%%%%%%%%%%%%%%%%%%%%%%%%%%%%%%%%%%%%%%%%%%%%%%%%%%%%%%%%
\begin{proof}
Note that from \Cref{assumption:lemma_assumption} and \Cref{lemma:lemma_3}
we have that $\vu^*$ and $\dvu$ coincide for the duplicate and non-duplicate problems.
Consequently, since $\nabla_\vq \ell$ and $\nabla_\vR \ell$ are only functions of $\vu^*$ and $\dvu$, these must
coincide.

Now, using the relationship between $\vd$ and $\bar{\vd}$ from
\eqref{eq:rel_d_dbar}, the definition of $\vM$ in \eqref{eq:M_def} implies that
\begin{align}
    \vM[(l-1)r+1:lr,\,l] &= 1\\
    \vM[(l-1)r+1:lr,\,b] &= 0
\end{align}
$\forall l,\,b \in \llbracket 1, \Nineqb \rrbracket$ and $j\neq l$. 
\par Hence,
\begin{align}
    (\vM\T \blambdadup)_l
    &= \sum_{k=1}^{Nr} \vM_{k, l} \lambdadup{k},\, \forall l \in \llbracket 1, \Nineqb \rrbracket \\
    &= \sum_{j=1}^{\xi_l} \lambdadup{\zeta_{l,j}} \\
    &= \bnu_l
\end{align}
and thus $\bnu = \vM\T \blambdadup$.
We can similarly show that $\bchi = \vM\T \diag(\blambdadup) \dblambdadup$:
\begin{align}
    \Big( \vM\T \diag(&\blambdadup) \dblambdadup) \Big)_l
    \nonumber
    \\
    &= \sum_{k=1}^{Nr} \vM_{k, l} \lamdlamdup{k}, \, \forall l \in \llbracket 1, \Nineqb \rrbracket \\
    &= \sum_{j=1}^{\xi_l} \lamdlamdup{\zeta_{l,j}} \\
    &= \bchi_l.
\end{align}
Hence, for $\nabla_{\bar{\vd}} \ell$, applying the chain rule yields,
\begin{align}
    \nabla_{\bar{\vd}} \ell 
    &= (\nabla_{\bar{\vd}} \vd) (\nabla_{\vd} \ell) \\
    &= \vM\T \nabla_{\vd} \ell \\
    &= -\vM\T \diag( \blambdadup^*) \dblambdadup \\
    &= -\bchi \\
    &= -\bgamma
\end{align}
which coincides with \eqref{eq:grad_d}.
Similarly for $\nabla_{\bar{\vC}} \ell$,
% \begin{align}
%     \nabla_{\bar{\vC}} \ell 
%     &= (\nabla_{\bar{\vC}} \vC) (\nabla_{\vC} \ell) \\
%     &= \vM\T \nabla_{\vC} \ell \\
%     &= \vM\T \left( \blambdadup^* \dvu\T
%     + \diag(\blambdadup^*) \dblambdadup  {\vu^*}\T \right) \\
%     &= \underbrace{\vM\T \blambdadup^*}_{\bnu} \dvu\T
%     + \underbrace{\vM\T\diag(\blambdadup^*) \dblambdadup}_{\bchi} {\vu^*}\T \\
%     &= \bnu \dvu\T + \bchi {\vu^*}\T \\
%     &= \blambdanondup^* \dvu\T + \bgamma {\vu^*}\T
% \end{align}
Noting that $\bar{\vc}_i\T$ represents the $i^\text{th}$column of $\bar{\vC}$ and that $\bar{\vc}_i\T = \vM \vc_i\T$ we have,
\begin{align}
    \nabla_{\bar{\vc}_i\T} \ell 
    &= (\nabla_{\bar{\vc}_i\T} \vc_i\T) (\nabla_{\vc_i\T} \ell) \\
    &= \vM\T \nabla_{\vc_i\T} \ell
\end{align}
Now collecting these vectors into matrices we have,
\begin{align}
    \nabla_{\bar{\vC}} \ell 
    &= (\nabla_{\bar{\vC}} \vC) (\nabla_{\vC} \ell) \nonumber\\ 
    &= \vM\T \left( \blambdadup^* \dvu\T
    + \diag(\blambdadup^*) \dblambdadup  {\vu^*}\T \right) \\
    &= \underbrace{\vM\T \blambdadup^*}_{\bnu} \dvu\T
    + \underbrace{\vM\T\diag(\blambdadup^*) \dblambdadup}_{\bchi} {\vu^*}\T \\
    &= \bnu \dvu\T + \bchi {\vu^*}\T \\
    &= \blambdanondup^* \dvu\T + \bgamma {\vu^*}\T
\end{align}
which coincides with \eqref{eq:grad_C}.
\end{proof}
%%%%%%%%%%%%%%%%%%%%%%%%%%%%%%%%%%%%%%%%%%%%%%%%%%%%%%%%%%%%%%%%%%
% \input{lemmas/same_gradient_proof}

\section{Connection between the Merged-OSQP Problem and the RNDCP}
\label{sec:relation_between_decentralized_and_RNDCP_section}
Assuming that convergence (i.e., consensus) is achieved, we can ignore the consensus related constraints and restate the KKT conditions (and the KKT system) for the Merged-OSQP problem so as to establish a connection to the KKT conditions (and the KKT system) of the non-duplicate problem. 
\par The per-agent (or per neighborhood) objective function for the Merged-OSQP problem is given by, \begin{align*}
    \mathcal{H}_i(\tilde{\vu}_i) &= \frac{1}{2}\tilde{\vu}_i\T\tilde{\vR}_i\tilde{\vu}_i + \tilde{\vq}_i\T\tilde{\vu}_i\\
    \text{where, } \tilde{\vR}_i &= \mathrm{bdiag}(\vR_i,\{\mathbf{0_{m\times m}}\}_{j=1}^r) \text{ and, } \\\tilde{\vq}_i&=\Big[\vG_i\T\frac{\partial V}{\partial \vx_i};\{\mathbf{0_{m\times 1}}\}_{j=1}^r\Big].
\end{align*}
The definitions of $\tilde{\vR}_i$ and $\tilde{\vq}_i$ above imply that the objective functions of each local QP consider only the objectives of the respective ego agents. This allows us to restate the objective of the Merged-OSQP problem and equate it to that of the non-duplicate centralized problem. Thus, we have, 
\begin{equation}
    \sum_{i=1}^N\mathcal{H}_i(\tilde{\vu}_i) = \mathcal{H}(\vu)
    \label{eqn:merged_nondup_obj_equivalence}
\end{equation}
where $\mathcal{H}(\vu)$ corresponds to the objective of the non-duplicate (or the full centralized) problem. 
\par Next, assuming we have converged, a complementary slackness condition for the Merged-OSQP problem can be constructed. To do this we refer to the following equations from the OSQP paper \cite[Equation 9]{osqp}, but based on notation from our main paper,
\begin{align}
&{{\blambda_i}_{-}^*}\T(\hat{\vz}_i^* - {\hat{\vz}_i}^{lb}) = 0\label{eqn:osqp_condition_1}\\
&{\blambda_i}_{-}^* = \min(\blambda_i^*,\,0)\label{eqn:osqp_condition_2}
\end{align}
where, ${\hat{\vz}_i}^{lb}$ is the lower bound on $\hat{\vz}_i^*$. Now, based on the constraints established in the main paper (i.e. $\vA_i\tilde{\vu}_i\leq \vd_i$), we have, \begin{equation*}
    -\infty < \hat{\vz}_i^* \leq \vd_i, \quad \forall \,i \in [[1,N]]
\end{equation*}
Since the ${\hat{\vz}_i}^{lb} \rightarrow -\infty$, $(\hat{\vz}_i^* - {\hat{\vz}_i}^{lb})$ will tend to a vector of large positive numbers. Additionally, because ${\blambda_i}_{-}^*$ can only be non-positive (based on \eqref{eqn:osqp_condition_2}), ${\blambda_i}_{-}^*$ must be equal to $\mathbf{0}$ so that the inner-product condition \eqref{eqn:osqp_condition_1} is satisfied (because a linear combination of large positive numbers with negative coefficients can never be equal to zero). This establishes that the following lower bound, 
\begin{equation}
    \blambda^* \geq \mathbf{0}.\label{eqn:osqp_comp_slack_1}
\end{equation}
On convergence we have, $\vA_i\tilde{\vu}_i^*=\tilde{\vz}_i^*=\hat{\vz}_i^*$. The condition on the upper bound of $\blambda$ mentioned in the OSQP paper then becomes, \begin{equation}
    {{\blambda_i}_+^*}\T (\vA_i\tilde{\vu}_i^* - \vd_i) = 0.\label{eqn:osqp_condition_3}
\end{equation}
Now, there are two possibilities, 
\begin{enumerate}
    \item \textbf{Tight constraints}: Where we have, $\vA_i\tilde{\vu}_i^* = \vd_i$. In this case we have ${\blambda_i}_+^*\geq 0$ (based on \eqref{eqn:osqp_comp_slack_1}.
    \item \textbf{Loose constraints}: Where we have, $\vA_i\tilde{\vu}_i^* < \vd_i$. Thus, for \eqref{eqn:osqp_condition_3} to hold, ${\blambda_i}_+^*=\mathbf{0}$ (because the coefficients of a linear combination of negative numbers have to be zero so that the sum is equal to zero).
\end{enumerate}
Thus, when constraints are loose, the upper bound on $\blambda$ will be 0. This therefore implies, \begin{equation}
    \mathrm{diag}(\blambda_i^*)(\vA_i\tilde{\vu}_i^* - \vd_i) = \mathbf{0}.
    \label{eqn:osqp_comp_slack_2}
\end{equation}
The equations \eqref{eqn:osqp_comp_slack_1} and \eqref{eqn:osqp_comp_slack_2} are the complementary slackness conditions that the solution of the converged Merged-OSQP must satisfy. 
\par Next, we can combine the constraints of all neighborhoods ($\vA_i$) into a single matrix. This would precisely be the matrix $\vC$ constructed from the matrices $\vA_i$ in the duplicate problem defined in \Cref{sec:construction_C_matrix}. Thus, the complementary slackness condition satisfied by the converged solution of Merged-OSQP is given by,\begin{align}
    & \mathrm{diag}(\blambda_\text{mosqp}^*)(\vC \vu_\text{mosqp}^* - \vd) = \mathbf{0} \nonumber\\
    & \blambda_\text{mosqp}^* \geq \mathbf{0} \label{eqn:final_comp_slack_mosqp}
\end{align} where, $\blambda_\text{mosqp}^*=[\blambda_1^*;\,\blambda_2^*;\,\ldots;\,\blambda_{Nr}^*]$ and $\vu_\text{mosqp}^*=\big[\tilde{\vu}_{1,1}^*;\,\tilde{\vu}_{2,1}^*;\,\ldots;\,\tilde{\vu}_{N,1}^*\big]$ where $\tilde{\vu}_{i,1}^*$ is the first block of the output of the $i^{\text{th}}$ local QP ($\tilde{\vu}_i^*$). However, because the solution of Merged-OSQP must also satisfy $\tilde{\vu}_i^*=\tilde{\vg}_i$ (i.e. we have consensus), we have $\vu_\text{mosqp}^* = \vu_\text{non-dup}^*$. 
\par The Lagrangian that the converged solution satisfies can be written as, \begin{align*}
    \mathcal{L}_\text{mosqp} &= \sum_{i=1}^N \Big(\mathcal{H}_i(\tilde{\vu}_i^*) + \sum_{i=1}^N {\blambda_i^*}\T(\vA_i\tilde{\vu}_i^* - \vd_i) \Big) \\
    &= \mathcal{H}(\vu_\text{non-dup}^*) + {\blambda_\text{mosqp}^*}\T(\vC \vu_\text{non-dup}^* - \vd)
\end{align*}
where, as mentioned earlier, we have ignored the consensus related variables.
\par The KKT conditions are given by,
\begin{align*}
    \bepsilon_\text{mosqp} = \nabla_{\bdelta_\text{mosqp}} \mathcal{L}_\text{mosqp} &= \begin{bmatrix}
     \nabla_{\bdelta_\text{mosqp}} \mathcal{H}(\tilde{\vu}^*) + \vC\T\blambda_\text{mosqp}^* \\ \mathrm{diag}(\blambda_\text{mosqp}^*)(\vC \vu_\text{mosqp}^* - \vd)
    \end{bmatrix} = \mathbf{0}.
\end{align*}
Thus, we have, 
\begin{align*}
\bepsilon_\text{mosqp} &= \begin{bmatrix}
     \vR\vu_\text{mosqp}^* + \vC\T\blambda_\text{mosqp}^* \\ \mathrm{diag}(\blambda_\text{mosqp}^*)(\vC \vu_\text{mosqp}^* - \vd)
    \end{bmatrix} \\[0.1cm]
    &= \begin{bmatrix}
     \vR\vu_\text{non-dup}^* + \vC\T\blambda_\text{mosqp}^* \\ \mathrm{diag}(\blambda_\text{mosqp}^*)(\vC \vu_\text{non-dup}^* - \vd)
    \end{bmatrix} = \mathbf{0}
\end{align*}
where $\bdelta_\text{mosqp}=[\vu_\text{mosqp}^*;\blambda_\text{mosqp}^*]$.
The Jacobian of the KKT conditions is given by, 
\begin{equation}
   \nabla_{\bdelta_\text{mosqp}}\bepsilon_\text{mosqp} = \begin{bmatrix}
   \vR & \vC\T \\ \mathrm{diag}(\blambda_\text{mosqp}^*)\vC & \mathrm{diag}(\vC\vu_\text{non-dup}^*-\vd) 
   \end{bmatrix} 
\end{equation}
which yields the following KKT system with the Jacobian matrix transposed, 
\begin{equation}
\begin{bmatrix}
\vR & \vC \T\diag(\blambda_{\text{mosqp}}^*) \\ 
 \vC & \diag(\vC \vu_\text{non-dup}^* - \vd)
\end{bmatrix}
\begin{bmatrix}
\mathrm{d}\vu_\text{mosqp} \\ \mathrm{d}\blambda_{\text{mosqp}}
\end{bmatrix}
=
-
\begin{bmatrix}
\nabla_{\vu_\text{non-dup}^*} \ell \\ 
\mathbf{0}
\end{bmatrix}.
\end{equation}
The connection between the KKT system above that the Merged-OSQP solution satisfies and the KKT system that the non-duplicate problem satisfies can be similarly established as was done for the duplicate centralized problem by combining KKT variables corresponding to duplicate constraints. This justifies the use of Lagrange multipliers obtained as a solution of Merged-OSQP with a duplicate centralized system for computation of backward pass gradients. So long as the LICQ is satisfied by the non-duplicate problem, the sum of the Lagrange multipliers will be unique.

\section{Derivation of Non-Linear Feynman-Kac}
\label{sec:nonlinear_feynman_kac_derivation_section}
For this derivation we will consider a simple stochastic optimal control problem for a control-affine system without any safety constraints. The corresponding HJB-PDE is given by, \begin{align*}
    &V_t + \inf_\mathbf{u} \Big\{\frac{1}{2}\text{tr}(V_{\mathbf{x} \mathbf{x}}\Sigma \Sigma^\mathrm{T})+V_\mathbf{x}^\mathrm{T}(\mathbf{f}+\mathbf{G} \mathbf{u})\\&\qquad\qquad\qquad\qquad+\sum_{i=1}^N c_i
        +\frac{1}{2}\mathbf{u}^\mathrm{T}\mathbf{R} \mathbf{u}\Big\}=0\\ &V(\mathbf{x}(T),T)=\phi(\mathbf{x}(T))
\end{align*}
where the subscripts $\vx$ and $t$ imply gradients with respect to $\vx$ and derivative with respect to time respectively. 
\par By setting the derivative of $\mathcal{H}$ with respect to $\vu$ equal to zero, we can derive the following closed-form expression for the optimal control, \begin{equation*}
    \mathbf{u}^*=-\mathbf{R}^{-1}\mathbf{G}^\mathrm{T} V_\mathbf{x}
\end{equation*}
Next we substitute for the optimal control into the HJB-PDE which allows us to drop the infimum operator giving, \begin{align*}
    V_t + \frac{1}{2}\tr(&V_{\mathbf{x} \mathbf{x}}\Sigma\Sigma^\mathrm{T}) + V_\mathbf{x}^\mathrm{T} \mathbf{f} \\&\quad+ \sum_{i=1}^N c_i - \frac{1}{2}V_\mathbf{x}^\mathrm{T}\mathbf{G} \mathbf{R}^{-1} \mathbf{G}^\mathrm{T}V_\mathbf{x}=0 
\end{align*}
Rearranging the above equation, we get the following expression for the derivative of $V$ w.r.t time $t$, \begin{align*}
    V_t = - \frac{1}{2}\tr(&V_{\mathbf{x} \mathbf{x}}\Sigma\Sigma^\mathrm{T}) - V_\mathbf{x}^\mathrm{T} \mathbf{f} \\&\quad- \sum_{i=1}^N c_i + \frac{1}{2}V_\mathbf{x}^\mathrm{T}\mathbf{G} \mathbf{R}^{-1} \mathbf{G}^\mathrm{T}V_\mathbf{x}
\end{align*}
Now, we apply Ito's formula to the value-function to obtain, \begin{align*}
    \mathrm{d}V &= V_t \mathrm{d}t + V_\mathbf{x}^\mathrm{T} \mathrm{d}\mathbf{x} + \frac{1}{2}\tr\big(V_{\mathbf{x}\mathbf{x}} \Sigma\Sigma^\mathrm{T}\big) \mathrm{d}t\\
    \mathrm{d}V &= \Big(- \frac{1}{2}\tr(V_{\mathbf{x} \mathbf{x}}\Sigma\Sigma^\mathrm{T}) - V_\mathbf{x}^\mathrm{T} \mathbf{f} \\&\quad\quad-\sum_{i=1}^N c_i + \frac{1}{2}V_\mathbf{x}^\mathrm{T}\mathbf{G} \mathbf{R}^{-1} \mathbf{G}^\mathrm{T}V_\mathbf{x} \Big) \mathrm{d}t \\&\quad\quad+ V_\mathbf{x}^\mathrm{T} \mathrm{d}\mathbf{x} + \frac{1}{2}\tr\big(V_{\mathbf{x}\mathbf{x}} \Sigma\Sigma^\mathrm{T}\big) \mathrm{d}t  \\
    &= \Big(- \frac{1}{2}\tr(V_{\mathbf{x} \mathbf{x}}\Sigma\Sigma^\mathrm{T}) - V_\mathbf{x}^\mathrm{T} \mathbf{f} \\
    &\quad\quad-\sum_{i=1}^N c_i + \frac{1}{2}V_\mathbf{x}^\mathrm{T}\mathbf{G} \mathbf{R}^{-1} \mathbf{G}^\mathrm{T}V_\mathbf{x} \Big) \mathrm{d}t \\
    &\quad\quad+ V_\mathbf{x}^\mathrm{T} \Big(\underbrace{(\mathbf{f} + \mathbf{G} \mathbf{u}^* )\mathrm{d}t + \Sigma\mathrm{d}\mathbf{w}}_{\mathrm{d}\mathbf{x}} \Big) + \frac{1}{2}\tr\big(V_{\mathbf{x}\mathbf{x}} \Sigma\Sigma^\mathrm{T}\big) \mathrm{d}t
\end{align*}
\begin{align*}
     &= {\color{red}\cancel{- \frac{1}{2}\tr(V_{\mathbf{x} \mathbf{x}}\Sigma\Sigma^\mathrm{T})}\mathrm{d}t} - {\color{ForestGreen}\cancel{V_\mathbf{x}^\mathrm{T} \mathbf{f}\,\mathrm{d}t}} \\& \quad \quad - \Big(\sum_{i=1}^N c_i - \frac{1}{2}V_\mathbf{x}^\mathrm{T}\mathbf{G} \mathbf{R}^{-1} \mathbf{G}^\mathrm{T}V_\mathbf{x}\Big)\mathrm{d}t + {\color{ForestGreen}\cancel{V_\mathbf{x}^\mathrm{T} \mathbf{f}\,\mathrm{d}t}} \\
    & \quad \quad + V_\mathbf{x}^\mathrm{T}(\mathbf{G} \mathbf{u}^* \mathrm{d}t + \Sigma\mathrm{d}\mathbf{w}) + {\color{red}\cancel{ \frac{1}{2}\tr\big(V_{\mathbf{x}\mathbf{x}} \Sigma\Sigma^\mathrm{T}\big) \mathrm{d}t}} \\
    &= -\Big(\sum_{i=1}^N c_i - \frac{1}{2}V_\mathbf{x}^\mathrm{T}\mathbf{G} \mathbf{R}^{-1} \mathbf{G}^\mathrm{T}V_\mathbf{x}\Big)\mathrm{d}t \\& \quad \quad +  V_\mathbf{x}^\mathrm{T}(\mathbf{G} \mathbf{u}^* \mathrm{d}t + \Sigma\mathrm{d}\mathbf{w}), \quad (\text{cancelling out common terms})\\
    &=-\Big(\sum_{i=1}^N c_i - \frac{1}{2}V_\mathbf{x}^\mathrm{T}\mathbf{G} \mathbf{R}^{-1} \mathbf{G}^\mathrm{T}V_\mathbf{x}\Big)\mathrm{d}t \\
    & \quad \quad +  V_\mathbf{x}^\mathrm{T}\mathbf{G} \Big(\underbrace{-\mathbf{R}^{-1}\mathbf{G}^\mathrm{T}V_\mathbf{x}}_{\mathbf{u}^*}\Big) \mathrm{d}t + V_\mathbf{x}^\mathrm{T}\Sigma\mathrm{d}\mathbf{w} \\
    &\mathrm{d}V = -\Big(\underbrace{\sum_{i=1}^N c_i + \frac{1}{2}V_\mathbf{x}^\mathrm{T}\mathbf{G} \mathbf{R}^{-1} \mathbf{G}^\mathrm{T}V_\mathbf{x}}_{\textrm{running  cost}}\Big)\mathrm{d}t + V_\mathbf{x}^\mathrm{T}\Sigma\mathrm{d}\mathbf{w}
    \end{align*}
This final expression gives us the required BSDE of the FBSDE system. The FSDE of the FBSDE system is the dynamics given by, 
\begin{equation*}
    \mathrm{d}\mathbf{x}(t) = \Big(\mathbf{f}\big(\mathbf{x}(t)\big) + \mathbf{G}\big(\mathbf{x}(t)\big)\mathbf{u}(t) \Big)\, \mathrm{d} t+ \Sigma\big(\mathbf{x}(t)\big)\, \mathrm{d}\mathbf{w}(t)
\end{equation*}
Putting everything togther, the Non-linear Feynman Kac theorem connects the unique solution of the HJB-PDE with the following system of FBSDEs,
\begin{align*}
    &\mathrm{d}\mathbf{x} = \Big(\mathbf{f}\big(\mathbf{x}\big) + \mathbf{G}\big(\mathbf{x}\big)\mathbf{u} \Big)\, \mathrm{d} t+ \Sigma\big(\mathbf{x}\big)\, \mathrm{d}\mathbf{w}, \quad \mathbf{x}(0)=\mathbf{x}_0\\
    &\mathrm{d}V = -\Big(\sum_{i=1}^N c_i + \frac{1}{2}V_\mathbf{x}^\mathrm{T}\mathbf{G} \mathbf{R}^{-1} \mathbf{G}^\mathrm{T}V_\mathbf{x}\Big)\mathrm{d}t + V_\mathbf{x}^\mathrm{T}\Sigma\mathrm{d}\mathbf{w},\\
    &V(\mathbf{x}(T),T)=\phi(\mathbf{x}(T)).
\end{align*}
%% Use plainnat to work nicely with natbib. 

\section{Merged-CADMM Method Additional Expressions} \label{sec: admm residuals}
The norms of the primal and dual residuals used in the Merged-CADMM termination criteria are given by:
\begin{align*}
r_{pri,1} & = \max_{i \in \llbracket 1, N \rrbracket } \| \vA_i \tilde{\vu}_i - \tilde{\vz}_i \|_{\infty}
\\ 
r_{pri,2} & = \max_{i \in \llbracket 1, N \rrbracket } \| \tilde{\vu}_i - \tilde{\vg}_i \|_{\infty}, 
\\
r_{dual,1} & = 
\max_{i \in \llbracket 1, N \rrbracket } \| \tilde{\vR}_i \tilde{\vu}_i + \tilde{\vq}_i + \vA_i \T \vy_i \|_{\infty}, \ 
\\ 
r_{dual,2} & = \max_{i \in \llbracket 1, N \rrbracket } \| \rho_2 (\tilde{\vg}_i^l - \tilde{\vg}_i^{l-1}) \|_{\infty},
\end{align*}
The corresponding thresholds are the following:
\begin{equation*}
\epsilon_{pri,a} = \epsilon_{\text{abs}} + \epsilon_{\text{rel}} 
\kappa_{pri,a} 
, \ 
\epsilon_{dual,a} = \epsilon_{\text{abs}} + \epsilon_{\text{rel}} 
\kappa_{dual,a}, \ a = 1,2
\end{equation*}
where:
\begin{align*}
\kappa_{pri,1} & = 
\max_{i \in \llbracket 1, N \rrbracket } \max{ 
\{ \| \vA_i \tilde{\vu}_i \|_{\infty}, 
\| \tilde{\vz}_i \|_{\infty} \}
},
\\
\kappa_{pri,2} & = 
\max_{i \in \llbracket 1, N \rrbracket } \max{ 
\{ \| \tilde{\vu}_i \|_{\infty}, 
\| \tilde{\vg}_i \|_{\infty} \}
},
\\
\kappa_{dual,1} & = 
\max_{i \in \llbracket 1, N \rrbracket } \max{ 
\{ \| \tilde{\vR}_i \tilde{\vu}_i \|_{\infty}, 
\| \tilde{\vq}_i \|_{\infty}, 
\| \vA_i \T \vy_i \|_{\infty} \}
},
\\
\kappa_{dual,2} & = \max_{i \in \llbracket 1, N \rrbracket } \| \bzeta_i \|_{\infty}.
\end{align*}

\section{Failure Probabilities for SCBF}
\label{sec:failure_bounds_section}
Restating the result from \cite[Proposition 1]{santoyo2021barrier} here for convenience of the reader: \textit{suppose there exists a twice differentiable function $B(\vx)$, that satisfies the following inequalities, \begin{align}
    &B(\vx)  \geq 0 \quad \forall\,\vx \in \mathbb{R}^{Nn_i} \\
    &B(\vx) \geq 1 \quad\forall\, \vx \in (\mathbb{R}^{Nn_i}\backslash \mathcal{S}) \\
    &\frac{\partial B}{\partial \vx}\T\Big(\vf + \vG \vu\Big) + \frac{1}{2}\text{tr}\bigg(\frac{\partial^2 B}{\partial \vx^2}\vSigma\vSigma\T\bigg) \leq -\alpha B(\vx) + \beta, \label{eqn:sm:safety_constraint}
\end{align}
where \eqref{eqn:sm:safety_constraint} is satisfied $\forall \,t\in[0,\,T]$ and $\forall\,\vx\in \mathbb{R}^{Nn_i}$ for some $\alpha \geq 0$ and $\beta \geq 0$. Define, \begin{align*}
    \rho_u &= \mathbb{P}\big\{\vx(t) \in (\mathbb{R}^{Nn_i}\backslash \mathcal{S}) \text{ for some } t \in [0,\,T]\big\}, \text{ and,}\\
    \rho_B &= \mathbb{P}\Big\{\sup_{t\in[0,\,T]}B(\vx)\geq1\Big\}
\end{align*}then, we have the following bounds,\begin{itemize}
    \item if $\alpha>0$ and $\beta\leq\alpha,\quad\rho_u\leq\rho_B\leq1-(1-B_0)e^{-\beta T}$ 
    \item if $\alpha>0$ and $\beta\geq\alpha,\quad\rho_u\leq\rho_B\leq \frac{B_0 + ( e^{\beta T}-1)\frac{\beta}{\alpha}}{e^{\beta T}}$ 
    \item if $\alpha=0,\quad\rho_u\leq\rho_B\leq B_0 + \beta T$
\end{itemize}
where $B_0=B\big(\vx(0)\big)$.}

\end{document}